\documentclass[10pt]{IEEETran}

\def \be {\begin{equation}}
\def \ee {\end{equation}}

\def \nn {\nonumber}

\usepackage{ifpdf}

\usepackage[dvips]{graphicx}
\usepackage{color}

\usepackage[cmex10]{amsmath}
\usepackage{amssymb}

\begin{document}
%
\title{\LARGE \bf Towards Manipulability of Interactive Lagrangian Systems}
%
%
%

\author{Hanlei~Wang 
\thanks{The author is with the Science and Technology on Space Intelligent Control Laboratory,
Beijing Institute of Control Engineering,
Beijing 100094, China (e-mail: hlwang.bice@gmail.com).
}
}
\maketitle

\begin{abstract}
This paper investigates manipulability of interactive Lagrangian systems with parametric uncertainty and communication/sensing constraints. Two standard examples are teleoperation with a master-slave system and teaching operation of robots. We here systematically formulate the concept of infinite manipulability for general dynamical systems, and investigate how such a unified motivation yields a design paradigm towards guaranteeing the infinite manipulability of interactive dynamical systems and in particular facilitates the design and analysis of nonlinear adaptive controllers for interactive Lagrangian systems. Specifically, based on a new class of dynamic feedback, we propose adaptive controllers that achieve both the infinite manipulability of the controlled Lagrangian systems and the robustness with respect to the communication/sensing constraints, mainly owing to the resultant dynamic-cascade framework. The proposed paradigm yields the desirable balance
between network coupling requirements and controlled dynamics
of human-system interaction. We also show that a special case of our main result resolves the longstanding nonlinear bilateral teleoperation problem with arbitrary unknown time-varying delay. Simulation results show the performance of the interactive robotic systems under the proposed adaptive controllers.
\end{abstract}


\section{Introduction}

One important trend of modern automatic machines is to facilitate the human-machine interaction. For instance, collaborative robots (the study of which has become particularly active in the robotics industry) are expected to be used in the scenario that the collaboration between the robots and human operators is frequently involved (e.g., the teaching operation in the standard ``teach-by-showing'' approach \cite{Weiss1987_JRA,Asada1989_TRA,Ikeuchi1994_TRA,Craig2005_Book}). Another example is teleoperation with a master-slave system in which case the slave robot is kept to be synchronized with the master robot that is guided by a human operator (see, e.g., \cite{Sheridan1993_TRA,Hokayem2006_AUT}). The fundamental issues behind these typical application scenarios are quite different from the common automatic control systems that emphasize stability with respect to an equilibrium; for instance, it is well known that the equilibrium of a teleoperator system is implicitly specified by the human operator (typically unknown a priori), and that it is the similar case for a robot manipulator under the standard teaching operation.

Historically, the control problems involved in teleoperation have received sustaining interest, which yields many significant results (see, e.g., the pioneering result in \cite{Anderson1989_TAC}, and \cite{Sheridan1993_TRA,Hokayem2006_AUT}). But the connection between teleoperation and standard control theory might be still relatively weak, mainly due to the lack of fundamental concepts that may enhance this connection, though there are some exceptional ones. In particular, the exploitation of the passivity concept in bilateral teleoperation (see, e.g., \cite{Anderson1989_TAC,Niemeyer1991_JOE}) is, in certain sense, a constructive attempt to address the connection issue (for instance, passivity often implies the potential stability of the system \cite{Anderson1989_TAC,Niemeyer1991_JOE,Niemeyer2004_IJRR}), and the past decades have witnessed the wide applications of this concept in teleoperation (see, e.g., \cite{Niemeyer2004_IJRR,Hokayem2006_AUT,Nuno2011a_AUT}). In recent years, benefiting from the extensive interest in control of multi-agent systems, many synchronization-based controllers have been proposed for teleoperators with their nonlinear dynamics being taken into account (see, e.g., \cite{Lee2006_TRO,Chopra2008_AUT,Nuno2009_IJRR,Liu2013_AUT,Wang2018_arXiv}) and the special case of the results in \cite{Nuno2011_TAC,Wang2014_TAC,Abdessameud2014_TAC} (focusing on consensus of networked Lagrangian systems on directed topologies) can also be used in a teleopertor system. 
A critical issue that spans the long history of teleoperation is the robustness with respect to the communication delay (see, e.g., \cite{Hokayem2006_AUT}), especially if the delay is time-varying. Many results (e.g., \cite{Anderson1989_TAC,Chopra2008_AUT,Nuno2010_AUT,Liu2013_AUT}) achieve robustness with respect to arbitrary unknown constant delay (which can also be referred to as delay-independent), yet this becomes frustrating as the delay is time-varying and in fact delay-independent result has not yet been witnessed in the case of time-varying delay. For instance, the results in \cite{Nuno2009_IJRR,Nuno2017_IJACSP} rely on designing the damping gain based on the upper bound of the time-varying delay and the result in \cite{Abdessameud2014_TAC} also exploits some a priori information of the delay for specifying the controller gains. 

For the more complicated networked Lagrangian systems, the issue associated with the coupling between the dynamics of each system and network interaction remains and is even much more severe, due to the fact that the topology might be directed and/or switching (see, e.g., \cite{Nuno2011_TAC,Wang2014_TAC,Liu2014_SCL,Liu2015_JFI}). For linear identical integrator systems or those systems that can be transformed to integrator systems by feedback, some strong results have been achieved for the consensus/synchonization problem---see, e.g., \cite{Chopra2008,Munz2011b_TAC}; in particular, both the time-varying delay and switching directed topologies are considered in \cite{Munz2011b_TAC} in the context of multiple identical single-integrator systems. The consideration of uncertain high-order systems on undirected jointly-connected topologies, using dynamic feedback, occurs in \cite{Rezaee2017_IJRNC}. The results for uncertain Lagrangian systems with switching topologies (and time delays) are presented in, e.g., \cite{Liu2014_SCL,Liu2014_JFI,Liu2015_JFI} and due to the use of static feedback, these results generally impose relatively strong requirement concerning the interaction topologies or time delays (for instance, the interaction topologies are required to be balanced or regular). Some attempts based on dynamic feedback design for uncertain Lagrangian systems occur in \cite{Wang2017_CAC} and \cite{Wang2017_AUTSubmitted}, which are mainly for realizing consensus of multiple Lagrangian systems with general switching directed topologies (and arbitrary time-varying communication delays), and the obtained topology/delay-independent solutions are mainly attributed to the dynamic feedback design. However, most of these results do not systematically consider the interaction between the network and an external subject (for instance, a human operator) while this becomes typical in the previously discussed problems of teleoperation and teaching operation.

The systems involving interaction with an external subject are typically referred to as interactive systems and in the specific context here as interactive Lagrangian systems. The investigation of such systems over the past mainly concentrates on the stability of the interactive systems; one common concept that is exploited is passivity since it is well recognized that passivity of the system typically implies stability as the system interacts with a human operator. On the other hand, achieving passivity shows some potential limitations as handling the system uncertainty and tough circumstance of the communication channel (see, for instance, \cite{Nuno2010_AUT,Nuno2011_TAC,Chopra2003_ACC}). The attempts to resolve this issue along other perspectives (not based on passivity) occur in, e.g., \cite{Nuno2010_AUT,Nuno2011_TAC,Liu2013_AUT,Wang2014_TAC,Abdessameud2014_TAC,Nuno2017_IJACSP}. Yet the systematic and rigorous formulation concerning the interactive behaviors of the combined system (for instance, the networked system and human operator) is still rarely witnessed. The results in \cite{Liu2013_AUT,Nuno2017_IJACSP} either consider some specific dynamics of human operators or present some particular ad hoc discussions concerning the human-robot interaction, and no rigorous or systematic formulation is presented, especially concerning the general fundamental mechanism behind the human-system interaction (beyond the standard passivity concept that mainly focuses on stability concerning the human-system interaction).

In this paper, we systematically formulate the concept of infinite (dynamical) manipulability to rigorously quantify the interactive behavior of Lagrangian systems under an external input action (force or torque), and the concept here extends/generalizes the one introduced in the specific context of consensus of networked robotic systems in \cite{Wang2018_arXiv} to general dynamical systems with mathematically rigorous formulation. Motivated in part by the result in \cite{Wang2018_arXiv} concerning the importance of existence of an infinite gain from the external force/torque to consensus equilibrium increment, a design paradigm towards guaranteeing the infinite manipulability of general interactive dynamical systems (namely, the existence of an infinite gain from the external input action to the specified output) is formally proposed. Differing from the concept of passivity in the literature that mainly addresses the stability issue of a human-manipulator interactive system, the concept of infinite manipulability is mainly for addressing the required amount of effort associated with the dynamic maneuvering of interactive (Lagrangian) systems. Specifically, based on a new class of dynamic feedback, we develop adaptive controllers to systematically address the issue of manipulability of a single Lagrangian system and that of networked Lagrangian systems with switching topology and unknown time-varying communication delays; the resultant closed-loop system is a dynamic-cascade one, which is in contrast to the system in \cite{Wang2017_AUTSubmitted} and also to the standard cascade system. The new feature
of the proposed adaptive controllers lies in the dynamic
feedback design of the reference velocity and acceleration
while the basic adaptive structure is the same as the standard one in \cite{Slotine1987_IJRR}. Our result covers two practically important applications, i.e., teaching operation of a robot manipulator and bilateral teleoperation with unknown time-varying communication delay. In particular, we demonstrate how the motivation of studying the manipulability of bilateral teleoperators leads to the first delay-independent solution to the longstanding benchmark nonlinear bilateral teleoperation problem with arbitrary unknown time-varying delay (to the best of our knowledge).

\section{Preliminaries}

\subsection{Equations of Motion of Lagrangian Systems}

The equations of motion of a $m$-DOF (degree-of-freedom) Lagrangian system can be written as \cite{Slotine1991_Book,Spong2006_Book}
\be
\label{eq:1}
M(q)\ddot q+C(q,\dot q)\dot q+g(q)=\tau
\ee
where $q\in R^m$ is the generalized position (or configuration), $M(q)\in R^{m\times m}$ is the inertia matrix, $C(q,\dot q)\in R^{m\times m}$ is the Coriolis and centrifugal matrix, $g(q)\in R^m$ is the gravitational torque, and $\tau\in R^m$ is the exerted control torque. Three typical properties concerning the dynamics (\ref{eq:1}) are listed as follows.

\emph{Property 1 (\cite{Slotine1991_Book,Spong2006_Book}):} The inertia matrix $M(q)$ is symmetric and uniformly positive definite.

\emph{Property 2 (\cite{Slotine1991_Book,Spong2006_Book}):} The Coriolis and centrifugal matrix $C(q,\dot q)$ can be appropriately chosen so that the matrix $\dot M(q)-2C(q,\dot q)$ is skew-symmetric.

\emph{Property 3 (\cite{Slotine1991_Book,Spong2006_Book}):} The dynamics (\ref{eq:1}) depends linearly on a constant dynamic parameter vector $\vartheta$ and this leads to
\be
M(q)\dot\zeta+C(q,\dot q)\zeta+g(q)=Y(q,\dot q,\zeta,\dot \zeta)\vartheta
\ee
where $\zeta\in R^m$ is a differentiable vector, $\dot \zeta$ is the derivative of $\zeta$, and $Y(q,\dot q,\zeta,\dot \zeta)$ is the regressor matrix.

\subsection{Input-Output/State Properties of Linear Time-Varying Systems}

The following lemmas concerning the input-output/state properties of linear time-varying systems are fundamental for most results given later.


\emph{Lemma 1 (\cite{Wang2017_AUTSubmitted}):} Consider a linear time-varying system with time-varying delays and with an external input
\begin{align}
\label{eq:4}
&\dot x={\mathcal F}_D (x)+ u\nn\\
&y=C^\ast(t) x
\end{align}
where $x$ is the state, $y$ is the output, ${\mathcal F}_D(\cdot)$ is a linear mapping with ${\mathcal F}_D(x)$ containing delayed version of $x$ and the time-varying delays are uniformly bounded, $C^\ast(t)$ is the output matrix and is uniformly bounded, and $u$ is the external input. Suppose that the output of the system (\ref{eq:4}) with $u=0$ uniformly exponentially converges to zero. Then
 \begin{enumerate}
 \item the system (\ref{eq:4}) is uniformly integral-bounded-input bounded-output stable, i.e., if  $\int_0^t u(\sigma)d\sigma\in{\mathcal L}_\infty$, then $y\in{\mathcal L}_\infty$;
\item if the time-varying delays $T_k(t)$, $k=1,2,\dots$ in ${\mathcal F}_D (x)$ satisfy the standard assumption that (see, e.g., \cite{Chopra2003_ACC,Niemeyer1998_ICRA})
 \be
 \label{eq:5}
 \dot T_k(t)\le 1-\epsilon
 \ee
 where $\epsilon$ is a constant satisfying the property that $0<\epsilon<1$, and if $\int_0^t u(\sigma)d\sigma+c\in{\mathcal L}_{p^\ast}$ with $c$ being certain constant vector, then $y\in{\mathcal L}_{p^\ast}$, ${p^\ast}\in[1,\infty)$.
\end{enumerate}


A special case of Lemma 1 (i.e., the case without involving time-varying delays) can be formulated as the following lemma.

\emph{Lemma 2:} Consider a linear time-varying system with an external input
\begin{align}
\label{eq:3}
&\dot x=A^\ast(t) x+ u\nn\\
&y=C^\ast(t) x
\end{align}
where $x$ is the state, $y$ is the output, $A^\ast(t)$ is the system coefficient matrix and is uniformly bounded, $C^\ast(t)$ is the output matrix and is uniformly bounded, and $u$ is the external input. Suppose that the output of the system (\ref{eq:3}) with $u=0$ uniformly exponentially converges to zero. Then
\begin{enumerate}
 \item the system (\ref{eq:3}) is uniformly integral-bounded-input bounded-output stable, i.e., if  $\int_0^t u(\sigma)d\sigma\in{\mathcal L}_\infty$, then $y\in{\mathcal L}_\infty$;

 \item if $\int_0^t u(\sigma)d\sigma+c\in{\mathcal L}_{p^\ast}$ with $c$ being certain constant vector, then $y\in{\mathcal L}_{p^\ast}$, ${p^\ast}\in[1,\infty)$.
\end{enumerate}

{\emph{Lemma 3 (\cite{Wang2017_AUTSubmitted}):} Consider a uniformly marginally stable linear time-varying system of the first kind (i.e., uniformly marginally stable linear systems with the state uniformly converging to a constant vector) with time-varying delays and an external input
\be
\label{eq:aa1}
\dot x={\mathcal F}_D(x)+u
\ee
where $x$ is the state, ${\mathcal F}_D(\cdot)$ is a linear mapping with ${\mathcal F}_D(x)$ containing delayed version of $x$ and the time-varying delays are uniformly bounded, and $u$ is the external input. Then
 \begin{enumerate}
 \item the system (\ref{eq:aa1}) is uniformly bounded-input differential-bounded-state stable, i.e., if $u\in{\mathcal L}_\infty$, then $\dot x\in{\mathcal L}_\infty$; if $u\in{\mathcal L}_{p^\ast}$, then $\dot x\in{\mathcal L}_{p^\ast}$, for $p^\ast\in[1,\infty)$; if $u\to 0$ as $t\to\infty$, then $\dot x\to 0$ as $t\to\infty$;
 \item the system (\ref{eq:aa1}) is uniformly integral-bounded-input bounded-${\mathcal F}_D(x)$ stable in the sense that if $\int_0^t u(\sigma)d\sigma\in{\mathcal L}_\infty$, then $\dot x-u={\mathcal F}_D(x)\in{\mathcal L}_\infty$.
 \end{enumerate}}

\section{Manipulability of Dynamical Systems}

We start by considering a standard simple example, namely, the motion of a point mass governed by
\be
m\ddot x=u+f
\ee
where $x\in R$ is the position of the point mass, $m\in R$ is the mass, $u\in R$ is the control input, and $f\in R$ is the external force from a subject (for instance, a human operator). Let us now consider the problem of the degree of the adjustability of the position $x$ under the action of the force $f$. Suppose that the control input $u$ takes the standard damping action as
\be
u=-b \dot x
\ee
with $b$ being a positive design constant, and we have that
\be
\label{eq:17}
m\ddot x=-b\dot x+f.
\ee
Since this is a linear time-invariant system, by following the standard practice, we obtain the transfer function from $f$ to $x$ as
\be
\label{eq:18}
G(p)=\frac{1}{m p^2+bp}
\ee
with $p$ denoting the Laplace variable, and further the ${\mathcal H}_\infty$ norm of $G(p)$ as
\begin{align}
\sup_{\omega}|G(j\omega)|=\sup_{\omega}\frac{1}{|\omega|\sqrt{m^2\omega^2+b^2}}=\infty
\end{align}
where $|\cdot|$ denotes the modulus of a complex number.
As is well recognized, the $\mathcal H_\infty$ norm (which is well known to be equal to the $\mathcal L_2$-gain for linear time-invariant systems) describes the energy-like relation between the input and output, i.e., the relation between the $\mathcal L_2$ norm of the output and that of the input. Specifically, for the example above, we have that (see, e.g., \cite{Desoer1975_Book,Ioannou1996_Book})
\be
\label{eq:aa2}
\sup_{f\ne 0}\frac{\|x\|_2}{||f||_2}=\sup_{\omega}|G(j\omega)|=\infty
\ee
where $\|\cdot\|_2$ denotes the standard $\mathcal L_2$ norm of a function. This would imply that an input with finite $\mathcal L_2$ norm holds the possibility of producing an output with infinite $\mathcal L_2$ norm, and consequently, it would be possible for a human operator to maneuver the position of the point mass to an arbitrary value with finite energy consumption (in the sense that the $\mathcal L_2$ norm of the exerted force is finite). This potentially reduces the required amount of effort from the human operator. In particular, consider an external force $f(t)=1/(t+1)$ [which is well known to have finite $\mathcal L_2$ norm (or be square-integrable)], and the output $x$, in accordance with (\ref{eq:aa2}), holds the possibility of having infinite $\mathcal L_2$ norm. The actual consequence can be illustrated by considering the output corresponding to this specific input, and by integrating (\ref{eq:17}) with respect time, it can be shown that [suppose that $x(0)=0$ and $\dot x(0)=0$]
\be
m\dot x=-b x+\ln(t+1)
\ee
and this implies that the output $x$ is the response of a standard stable filter with an unbounded input $\ln(t+1)$. It is well recognized that the output $x$ (i.e., the position of the point mass) converges to infinity as $t\to\infty$ ($x$ has infinite $\mathcal L_2$ norm), in comparison with the fact that the external force $f(t)=1/(t+1)$ is square-integrable and actually converges to zero as $t\to\infty$ [$f(t)$ has finite $\mathcal L_2$ norm].

We now formally introduce the concept of infinite manipulability or infinite dynamical manipulability for general dynamical systems, which generalizes the one introduced in the specific context of consensus of networked robotic systems in \cite{Wang2018_arXiv} to consider general dynamical systems with mathematically rigorous formulation. 
Manipulability of a dynamical system in terms of its specified output basically describes the degree of adjustability of the output corresponding to an external input action, and it is essentially equivalent to the standard concept of reachability/controllability or output reachability/controllability for dynamical systems. The distinguished or particular point concerning (dynamical) manipulability may lie in its emphasis on the physical interactive maneuvering behavior of (controlled) dynamical systems acted upon by an external subject, in contrast to the concept of reachability/controllability or output reachability/controllability that is typically associated with stability or stabilizability of a dynamical system itself. What is of particular interest, as is shown in our later result, is the infinite manipulability and it is typically associated with marginally stable dynamical systems.

\emph{Definition 1:}
\begin{enumerate}

\item A dynamical system is said to be infinitely manipulable if the gain of the mapping from the external input action
 to the output is infinite. 

\item A dynamical system is said to be infinitely manipulable with degree $k$, $k=1,2,\dots$ if the gain of the mapping from the external input action to the output is infinite and if the mapping contains $k$ pure integral operations with the infinite portion of the gain being solely due to the $k$ pure integral operations.
\end{enumerate}


  The ``gains'' used in the standard input-output analysis (see, e.g., \cite{Desoer1975_Book,Ioannou1996_Book,Schaft2000_Book}) can be directly adopted for quantifying the system (dynamical) manipulability (as also illustrated in the above simple example), and for facilitating the formulation later,  the quantification of (dynamical) manipulability of a dynamical system over the interval $[0,t]$ is denoted by $\mathcal M_{f\mapsto y}^t$ with $y$ denoting the output and $f$ the external input action, 
and $\mathcal M_{f\mapsto y}^\infty$ is typically denoted by $\mathcal M_{f\mapsto y}$ for conciseness.

In many applications, it is desirable to maintain the infinite manipulability of the system concerning the specific output (e.g., for reducing the amount of effort exerted by the human operator in the course of adjusting the system equilibrium). On the other hand, for a system with overly high manipulability, it might be difficult for a human operator to accurately adjust the system equilibrium.

For instance, the mapping given by (\ref{eq:18}) contains one pure integral operation [i.e., there is only one
 factor $1/p$ in $G(p)$], and thus the system (\ref{eq:17}) is said to be infinitely manipulable with degree one. In contrast, if the damping parameter $b$ is set to be zero, the manipulability degree of the system becomes two. It would typically be more difficult to manipulate an infinitely manipulable system with degree two in comparison with an infinitely manpulable system with degree one. Intuitively, we can consider the manipulation of a point mass on a frictionless horizontal plane and that of a point mass on a horizontal plane with viscous friction. The accurate positioning for the case without any friction is expected to be much more difficult than that for the case with viscous friction. In many practical applications, the infinite manipulability with degree one tends to be more feasible and safer.

We further discuss the case of using a time-varying and uniformly positive damping gain $b(t)$, and it is well known that the approach relying on calculating the $\mathcal H_\infty$ norm is no longer applicable for this case. As now standard, consider the Lyapunov function candidate
\be
V=\frac{1}{2}m \dot x^2
\ee
whose derivative along the trajectories of the system can be written as
\be
\dot V=-b\dot x^2+\dot x f.
\ee
By resorting to the standard basic inequalities, we have that
\be
\dot V\le -\frac{b}{2}\dot x^2+\frac{1}{2b}f^2
\ee
which directly yields
\be
V(t)+\int_0^t \frac{b(\sigma)}{2}\dot x^2(\sigma)d\sigma\le V(0)+\int_0^t \frac{1}{2b(\sigma)}f^2(\sigma)d\sigma
.\ee
By following the typical practice (see, e.g., \cite{Khalil2002_Book}), we obtain that the $\mathcal L_2$-gain of the mapping from $f$ to $\dot x$ is less than or equal to
$1/{\min\{b(t)\}}$ (i.e., finite). Therefore, the $\mathcal L_2$-gain of the mapping from $f$ to $x$ is the composite of a finite $\mathcal L_2$-gain (less than or equal to $1/{\min\{b(t)\}}$) and the $\mathcal L_2$-gain of a pure integral operation (which is well known to be $\sup_\omega\frac{1}{|\omega|}=\infty$). Hence the manipulability of the system is infinite with degree one.

For conciseness of the subsequent formulations and demonstrations, the manipulability analysis based on input-output gains in the sequel, if not particularly mentioned, follows the standard practice; see also the relevant results in, e.g., \cite{Vidyasagar1993_Book,Rotea1993_AUT,Schaft2000_Book,Khalil2002_Book,Desoer1975_Book,Ioannou1996_Book,Sontag1998_SCL,Angeli2000_TAC} for the details.

\section{Manipulability of A Lagrangian System}

For the Lagrangian system given by (\ref{eq:1}), we investigate the manipulability of the system with its generalized position or velocity as the output. We expect to realize the infinite manipulability of the system in terms of the output. In particular, the infinite manipulability of the system in terms of its generalized position has important applications in teaching operation of robot manipulators.

\subsection{Damping Control With Gravitational Torque Compensation}
Consider the standard damping control with the gravitational torque compensation
\be
\label{eq:24}
\tau=-\alpha\dot q+g(q)
\ee
where $\alpha$ is a positive design constant. Under an external input action $\tau_h$, this controller yields
\be
M(q)\ddot q+C(q,\dot q)\dot q=-\alpha\dot q+\tau_h
.\ee
The above system defines a mapping from $\tau_h$ to $q$, and as is previously discussed, the gain of this mapping quantifies the manipulability of the system. For analyzing the manipulability of the system, as is typically done, consider the Lyapunov function candidate
\be
V=\frac{1}{2}\dot q^T M(q)\dot q
\ee
and we then have that (using Property 2)
\be
\label{eq:27}
\dot V=-\alpha\dot q^T \dot q+\dot q^T\tau_h.
\ee
Using the following result derived from the standard basic inequalities
\be
\dot q^T\tau_h\le \frac{\alpha}{2}\dot q^T\dot q+\frac{1}{2\alpha}\tau_h^T \tau_h,
\ee
we obtain from (\ref{eq:27}) that
\be
\dot V\le-\frac{\alpha}{2}\dot q^T\dot q+\frac{1}{2\alpha}\tau_h^T \tau_h.
\ee
Then by following the typical practice (see, e.g., \cite{Khalil2002_Book}), we obtain that the $\mathcal L_2$-gain from $\tau_h$ to $\dot q$ is less than or equal to $1/\alpha$. On the other hand, the $\mathcal L_2$-gain from $\dot q$ to the position increment $q-q(0)$ is that of a pure integral operation, i.e., $\sup_{\omega}\frac{1}{|\omega|}=\infty$. Therefore, the $\mathcal L_2$-gain from $\tau_h$ to $q-q(0)$ satisfies the property that ${\mathcal M}_{\tau_h\mapsto q-q(0)}\le \frac{1}{\alpha}\sup_{\omega}\frac{1}{|\omega|}=\infty$ and in addition ${\mathcal M}_{\tau_h\mapsto q-q(0)}$ has the same order as the upper bound (as in the typical practice of calculating the gains for general nonlinear dynamical systems), and thus the infinite manipulability of the system with degree one is ensured. In addition, in the case that $\tau_h=0$ (i.e., the system is in free motion), we immediately obtain by the typical practice that $\dot q\to 0$ as $t\to\infty$.

\emph{Theorem 1:} The controller (\ref{eq:24}) for the Lagrangian system given by (\ref{eq:1}) ensures that the system with $q$ as the output is infinitely manipulable with degree one.

\emph{Remark 1:} The damping control with gravitational torque compensation as well as the stability of the closed-loop system in the case that $\tau_h=0$ is well recognized (especially in the standard teaching operation of robots), and the result here is for revisiting this standard problem in the context of rigorously analyzing the manipulability of the system and for showing that the system is actually infinitely manipulable with degree one.

\subsection{Adaptive Control}

In the presence of parametric uncertainty, the gravitational torque compensation is no longer accurate which would possibly result in the reduction of  manipulability of the system. More importantly, we expect to rigorously address the quantitative performance of the system (e.g., guaranteeing the efficiency of teaching operation of a robot manipulator) in addition to the manipulability even if we do not exactly know the system model or the system model is subjected to a variation. This can be accommodated in part by the flexibility provided by adaptive control (see, e.g., \cite{Slotine1989_Aut,Wang2017_TAC}).

We first introduce a vector $z\in R^m$ by
\be
\label{eq:29}
\dot z=-\alpha\dot q+\lambda_{\mathcal M}(\dot q-z)
\ee
where $\alpha$ and $\lambda_{\mathcal M}$ are positive design constants, and define
\be
\label{eq:30}
s=\dot q-z.
\ee
The adaptive controller is given as
\begin{align}
\label{eq:31}
\tau=&-K s+Y(q,\dot q,z,\dot z)\hat\vartheta\\
\label{eq:32}
\dot{\hat \vartheta}=&-\Gamma Y^T(q,\dot q,z,\dot z)s
\end{align}
where $K$ and $\Gamma$ are symmetric positive definite matrices, and $\hat\vartheta$ is the estimate of $\vartheta$. The dynamics of the system can then be described by
\be
\label{eq:33}
\begin{cases}
\ddot q=-\alpha\dot q+\lambda_{\mathcal M}s+\dot s\\
M(q)\dot s+C(q,\dot q)s=-K s+Y(q,\dot q,z,\dot z)\Delta \vartheta+\tau_h\\
\dot{\hat \vartheta}=-\Gamma Y^T(q,\dot q,z,\dot z)s
\end{cases}
\ee
where $\Delta \vartheta=\hat\vartheta-\vartheta$. Equation (\ref{eq:33}) defines a system which we refer to as dynamic-cascade system since the cascade component $\lambda_{\mathcal M}s+\dot s$ involves both the vector $s$ [generated by the lower two subsystems of (\ref{eq:33})] and its derivative $\dot s$, in contrast to the system in \cite{Wang2017_AUTSubmitted} and also to the standard cascade system.

\emph{Remark 2:} The insertion of the action $\lambda_{\mathcal M}(\dot q-z)$ in (\ref{eq:29}) is for ensuring the infinite  manipulability of the system in terms of its generalized position. The basic structure of the adaptive controller given by (\ref{eq:31}) and (\ref{eq:32}) follows the fundamental result in \cite{Slotine1987_IJRR} (this adaptive structure is also exploited in the sequel in the context of networked Lagrangian systems and teleoperator systems), and new reference velocity and acceleration [i.e., $z$ and $\dot z$ given by (\ref{eq:29})] are introduced to address the manipulability issue of a single Lagrangian system with parametric uncertainty.

As is standard (see, e.g., \cite{Slotine1987_IJRR,Ortega1989_AUT}), consider the Lyapunov-like function candidate
\be
\label{eq:34}
V=\frac{1}{2}s^T M(q)s+\frac{1}{2}\Delta\vartheta^T \Gamma^{-1}\Delta \vartheta
\ee
and its derivative along the trajectories of the system can be written as (using Property 2)
\be
\label{eq:35}
\dot V=-s^T Ks +s^T \tau_h.
\ee

\emph{Theorem 2: } The adaptive controller given by (\ref{eq:31}) and (\ref{eq:32}) for the Lagrangian system given by (\ref{eq:1}) ensures that $\dot q\to 0$ in the case that $\tau_h=0$. In addition, the system with $q$ as the output is infinitely manipulable with degree one.

\emph{Proof:}  In the case that $\tau_h=0$, from (\ref{eq:35}), we directly obtain that $\dot V=-s^TK s\le 0$, which implies that $s\in \mathcal L_2\cap \mathcal L_\infty$ and $\hat\vartheta\in\mathcal L_\infty$. This immediately yields the result that $\int_0^t \dot s(\sigma)d\sigma=s-s(0)\in\mathcal L_\infty$ and $\int_0^t \dot s(\sigma)d\sigma+s(0)=s\in{\mathcal L}_2$. Consider the first subsystem of (\ref{eq:33}) with $\dot q$ as the output, and in accordance with the standard linear system theory, the output can be considered as the superposition of the output under the input $\lambda_{\mathcal M}s$ and that under the input $\dot s$. The system given by $\ddot q=-\alpha\dot q$ with $\dot q$ as the output is exponentially stable and strictly proper by the standard linear system theory. Hence from the input-output properties of exponentially stable and strictly proper linear systems \cite[p.~59]{Desoer1975_Book}, we obtain that the output of the system corresponding to the input $\lambda_{\mathcal M}s$ is square-integrable and bounded. From Lemma 2, we obtain that the output corresponding to the input $\dot s$ is also square-integrable and bounded. This implies that $\dot q\in{\mathcal L}_2\cap{\mathcal L}_\infty$ in accordance with the standard superposition principle for linear systems. We then obtain that $z=\dot q-s\in{\mathcal L}_2\cap {\mathcal L}_\infty$, and this leads us to obtain from (\ref{eq:29}) that $\dot z\in{\mathcal L}_2\cap{\mathcal L}_\infty$. From the second subsystem of (\ref{eq:33}) and using Property 1, we obtain that $\dot s\in{\mathcal L}_\infty$. Therefore $\ddot q\in{\mathcal L_\infty}$ and thus $\dot q$ is uniformly continuous. From the properties of square-integrable and uniformly continuous functions \cite[p.~232]{Desoer1975_Book}, we obtain that $\dot q\to0$ as $t\to\infty$.

We now consider the manipulability of the system with $q$ as the output. First, consider the mapping from $\tau_h$ to $s$, and as is well recognized in adaptive control, this mapping is obviously not $\mathcal L_\infty$-gain bounded yet it is $\mathcal L_2$-gain bounded and also $\mathcal L_2$$\mapsto$$\mathcal L_\infty$-gain (see, e.g., \cite{Rotea1993_AUT,Sontag1998_SCL,Angeli2000_TAC} for the detail) bounded. In fact, by using the standard basic inequalities, we can obtain from (\ref{eq:35}) that
\be
\dot V\le -\frac{1}{2}s^T Ks+\frac{1}{2}\tau_h K^{-1}\tau_h
\ee
and this implies that the $\mathcal L_2$-gain from $\tau_h$ to $s$ is less than or equal to $1/{\lambda_{\min}\{K\}}$ with $\lambda_{\min}\{\cdot\}$ denoting the minimum eigenvalue of a matrix and that if $\tau_h\in{\mathcal L}_2$, then $s\in\mathcal L_2\cap\mathcal L_\infty$. From the first subsystem of (\ref{eq:33}), we can investigate the mapping from $s$ to $q$, and the output $\dot q$ of this subsystem, due to its linear nature, can be considered to be the superposition of two outputs corresponding to two inputs $s$ and $\dot s$, respectively in accordance with the standard superposition principle for linear
systems. From Lemma 2, the input $\dot s$ yields a square-integrable and bounded output since $\int_0^t \dot s(\sigma)d\sigma+s(0)=s\in{\mathcal L}_2$ and $\int_0^t s(\sigma)d\sigma=s-s(0)\in{\mathcal L}_\infty$ in the case that $s\in{\mathcal L}_2\cap{\mathcal L}_\infty$. In addition, the input $\dot s$ yields the first portion of $q$ (bounded), which can be observed by the integral operation of
$$\ddot q=-\alpha\dot q+\dot s$$
with respect to time. In fact, we have that
\be
\label{eq:37}
\dot q=-\alpha q+\alpha q(0)+\dot q(0)-s(0)+s
\ee
which means that $q$ in (\ref{eq:37}) is the output of an exponentially stable linear system with bounded input and is thus bounded by the standard linear system theory. The input $s$, on the other hand, only yields a square-integrable and bounded output (second portion of $\dot q$) and it does not lead to the boundedness of the second portion of $q$ since the integral operation of
$$
\ddot q=-\alpha\dot  q+\lambda_{\mathcal M} s
$$
with respect to time gives
\be
\label{eq:38}
\dot q=-\alpha q+\alpha q(0)+\dot q(0) +\lambda_{\mathcal M}\int_0^t s (\sigma) d\sigma
\ee
and $s\in{\mathcal L}_2\cap{\mathcal L}_\infty$\footnote{It is well known that square-integrability of a function does not imply that its integral is bounded and this function holds the possibility of being integral unbounded.}. We now explicitly calculate the $\mathcal L_2$-gain of the mapping from $s$ to $q$ and to this end, we combine (\ref{eq:37}) and (\ref{eq:38}) as
\be
\dot q=-\alpha [q-\psi_0]+\lambda_{\mathcal M}\int_0^t s(\sigma)d\sigma+s
\ee
with $\psi_0=[\alpha q(0)+\dot q(0)-s(0)]/\alpha$. By following the standard practice, the $\mathcal L_2$-gain from $\lambda_{\mathcal M}\int_0^t s(\sigma)d\sigma+s$ to $q-\psi_0$ can be derived as $1/\alpha$ (i.e., the $\mathcal H_\infty$ norm of the transfer function), and the $\mathcal L_2$-gain from $s$ to $\lambda_{\mathcal M}\int_0^t s(\sigma)d\sigma+s$ can be derived as $\sqrt{\lambda_{\mathcal M}^2\sup_{\omega} \frac{1}{|\omega|^2}+1}$. Therefore, the $\mathcal L_2$-gain from $\tau_h$ to $q-\psi_0$ satisfies
\be
{\mathcal M}_{\tau_h\mapsto q-\psi_0}\le
\frac{1}{\alpha \lambda_{\min}\{K\}}\sqrt{\lambda_{\mathcal M}^2\sup_{\omega} \frac{1}{|\omega|^2}+1}=\infty.
\ee
As in the typical practice of calculating the gains for general nonlinear dynamical systems, this shows that the system is infinitely manipulable with degree one. \hfill{\small {$\blacksquare $}}


\emph{Remark 3:} Following the result in \cite{Slotine1989_Aut}, we choose the matrix $K$ in (\ref{eq:31}) as $K=\lambda_c \hat M(q)$ and modify the regressor matrix in (\ref{eq:32}) as $Y^\ast=Y(q,\dot q,z,\dot z-\lambda_c s)$ where $\lambda_c$ is a positive design constant and $\hat M(q)$ is the estimate of $M(q)$ [which is obtained by replacing $\vartheta$ in $M(q)$ with $\hat\vartheta$]. This immediately yields the dynamics $M(q)\dot s+C(q,\dot q)s=-\lambda_c M(q)s+Y^\ast\Delta\vartheta+\tau_h$ [in comparison with the second subsystem of (\ref{eq:33})] and the result that $\dot V=-\lambda_c s^T M(q)s+s^T \tau_h$ (similar to \cite{Slotine1989_Aut}), upon which it can be demonstrated by following similar arguments as in \cite{Slotine1989_Aut,Wang2017_TAC} that this modification efficiently guarantees (improves) the performance (and also facilitates the quantification of the performance) concerning the dynamic response from $\tau_h$ to $s$ in the sense of certainty equivalence. Considering the fact that the dynamic response from $s$ to $q-\psi_0$ is described by a standard linear time-invariant system, the performance of the dynamic response from $\tau_h$ to $q-\psi_0$ is thus guaranteed/improved and also well quantifiable in the sense of certainty equivalence. 


\subsection{Velocity as the Output}

The controllers above ensure the infinite manipulability of the system with the generalized position as the output, and on the other hand, we note that the gain of the mapping from the external input action to the velocity $\dot q$ is actually finite, yielding finite manipulability of the system with respect to the velocity. This would imply that for adjusting the velocity of the system to a desired value, the external input has to hold a force or torque constantly. To achieve the infinite manipulability of the system in terms of the velocity, we can simply set $\alpha=0$ and then the gain of the mapping from $\tau_h$ to $\dot q$ becomes infinite and in addition this mapping contains one pure integral operation. Hence, the infinite manipulability with degree one can be guaranteed.

\section{Manipulability of Networked Lagrangian Systems}

In this section, we consider $n$ Lagrangian systems with the dynamics of the $i$-th system being governed by \cite{Slotine1991_Book,Spong2006_Book}
\be
\label{eq:41}
M_i(q_i)\ddot q_i+C_i(q_i,\dot q_i)\dot q_i+g_i(q_i)=\tau_i
\ee
where $q_i\in R^m$ is the generalized position or configuration, $M_i(q_i)\in R^{m\times m}$ is the inertia matrix, $C_i(q_i,\dot q_i)\in R^{m\times m}$ is the Coriolis and centrifugal matrix, $g_i(q_i)\in R^m$ is the gravitational torque, and $\tau_i\in R^m$ is the exerted control torque.

We briefly introduce the graph theory in the context involving $n$ Lagrangian systems by following \cite{Olfati-Saber2004_TAC,Ren2005_TAC,Ren2008_Book}.

As now standard, we adopt a directed graph ${\mathcal G}=({\mathcal V},{\mathcal E})$ for describing the interaction topology among the $n$ systems where ${\mathcal V}=\left\{1,\dots,n\right\}$ is the vertex set that denotes the collection of the $n$ systems and ${\mathcal E}\subseteq {\mathcal V}\times {\mathcal V}$ is the edge set that denotes the information interaction among the $n$ systems. A graph is said to contain a directed spanning tree if there exists a vertex $k^\ast\in{\mathcal V}$ so that any other vertex of the graph has a directed path to $k^\ast$, where the vertex $k^\ast$ is referred to as the root of the graph. Denote by $\mathcal{N}_i=\left\{j|(i,j)\in \mathcal{E}\right\}$ the set of neighbors of the $i$-th system. The weighted adjacency matrix $\mathcal W=[w_{ij}]$ associated with $\mathcal G$ is defined in accordance with the rule that $w_{ij}$ is strictly positive in the case that $j\in{\mathcal N}_i$, and $w_{ij}=0$ otherwise. The standard assumption regarding the diagonal entries of the matrix $\mathcal W$ that $w_{ii}=0$, $\forall i=1,\dots,n$ is adopted. With the definition of the weighted adjacency matrix $\mathcal W$, the Laplacian matrix $\mathcal L_w=[\ell_{w,ij}]$ associated with $\mathcal G$ is defined in accordance with the rule that $\ell_{w,ij}=\Sigma_{k=1}^n w_{ik}$ if $i=j$, and $\ell_{w,ij}=-w_{ij}$ otherwise.

In the case that the interaction topology switches, the interaction graph among the systems becomes time-varying. Denote by ${\mathcal G}_S=\{\mathcal G_1,\dots,\mathcal G_{n_s}\}$ the set of the interaction graphs among the $n$ systems, and these graphs share the same vertex set $\mathcal V$ yet their edge sets are typically different. The union of a collection of graphs $G_{i_1}, \dots,G_{i_s}$ with $i_s\le n_s$ is a graph with the vertex set given by $\mathcal V$ and the edge set given by the union of the edge sets of $G_{i_1}, \dots,G_{i_s}$.  Denote by $t_0,t_1,t_2,\dots$ a series of time instants at which the interaction graph switches, and it is assumed that these instants satisfy the standard property that $0=t_0<t_1<t_2<\dots$ and that $T_D\le t_{k+1}-t_k<T_0$, $\forall k=0,1,\dots$ with $T_D$ and $T_0$ being two positive constants where the assumption concerning the dwell time that $t_{k+1}-t_k\ge T_D$, $\forall k=0,1,\dots$ is standard for the case of switching interaction topology (refer \cite{Ren2008_Book} for the details).


In the following, we design adaptive controllers to realize consensus of the $n$ Lagrangian systems with switching directed topologies (and unknown time-varying communication delay) and simultaneously ensure the infinite manipulability of the system. 

\subsection{Consensus With Switching Topology}

We first consider the case of $n$ Lagrangian systems with switching directed topologies. Define a vector $z_i\in R^m$ by the following dynamic system
\begin{align}
\label{eq:42}
\dot z_i=&-{\alpha \dot q_i}+\lambda_{\mathcal M}(\dot q_i-z_i)\nn\\
&-\Sigma_{j\in{\mathcal N}_i(t)}w_{ij}(t)[(\dot q_i+\alpha q_i)-(\dot q_j+\alpha q_j)]
\end{align}
with $\alpha$ and $\lambda_{\mathcal M}$ being positive design constants, and define a sliding vector
\be
\label{eq:43}
s_i=\dot q_i-z_i.
\ee
The adaptive controller is given as
\begin{align}
\label{eq:44}
\tau_i=&-K_i s_i+Y_i(q_i,\dot q_i,z_i,\dot z_i)\hat\vartheta_i\\
\label{eq:45}
\dot{\hat \vartheta}_i=&-\Gamma_i Y_i^T(q_i,\dot q_i,z_i,\dot z_i)s_i
\end{align}
where $K_i$ and $\Gamma_i$ are symmetric positive definite matrices, $\hat\vartheta_i$ is the estimate of the unknown parameter vector $\vartheta_i$, and the regressor matrix $Y_i(q_i,\dot q_i,z_i,\dot z_i)$ and the unknown parameter vector $\vartheta_i$ are defined in accordance with the standard linearity-in-parameter property of the Lagrangian system (see, for instance, \cite{Slotine1991_Book,Spong2006_Book}), i.e.,
\be
M_i(q_i)\dot z_i+C_i(q_i,\dot q_i)z_i+g_i(q_i)=Y_i(q_i,\dot q_i,z_i,\dot z_i)\vartheta_i.
\ee The dynamics of the $i$-th system can then be described by
\be
\label{eq:46}
\begin{cases}
\ddot q_i=-{\alpha \dot q_i}-\Sigma_{j\in{\mathcal N}_i(t)}w_{ij}(t)[(\dot q_i+\alpha q_i)-(\dot q_j+\alpha q_j)]\\
\quad\quad+\lambda_{\mathcal M} s_i+\dot s_i\\
M_i(q_i)\dot s_i+C_i(q_i,\dot q_i)s_i=-K_i s_i+Y_i(q_i,\dot q_i,z_i,\dot z_i)\Delta \vartheta_i\\
\dot{\hat \vartheta}_i=-\Gamma_i Y_i^T(q_i,\dot q_i,z_i,\dot z_i)s_i
\end{cases}
\ee
where $\Delta \vartheta_i=\hat\vartheta_i-\vartheta_i$. The above system, similar as before, is also a dynamic-cascade system
in the sense that the cascade component $\lambda_{\mathcal M} s_i+\dot s_i$ involves both the vector $s_i$ and its derivative $\dot s_i$. 

\emph{Theorem 3:} Suppose that there exist an infinite number of uniformly bounded intervals $[t_{i_\ell},t_{i_{\ell+1}})$, $\ell=1,2,\dots$ with $t_{i_1}=t_0$ satisfying the property that the union of the interaction graphs in each interval contains a directed spanning tree. Then the adaptive controller given by (\ref{eq:44}) and (\ref{eq:45}) with $z_i$ being given by (\ref{eq:42}) ensures 1) the consensus of the $n$ systems in free motion (i.e., no external physical interaction), i.e., $q_i-q_j\to 0$ and $\dot q_i\to 0$ as $t\to \infty$, $\forall i,j=1,\dots,n$ and 2) the infinite manipulability of the system with degree one in terms of an external physical input action at the torque level and the consensus equilibrium increment.

\emph{Proof:} We first follow the standard practice to analyze the lower two subsystems of (\ref{eq:46}) (see, e.g., \cite{Ortega1989_AUT,Slotine1987_IJRR}). Specifically, consider the  Lyapunov-like function candidate $
V_i=(1/2)s_i^T M_i(q_i)s_i+(1/2)\Delta \vartheta_i^T\Gamma_i^{-1}\Delta\vartheta_i
$ and its derivative along the trajectories of the system can be written as
$
\dot V_i=-s_i^T K_i s_i\le 0
$ with the skew-symmetry of $\dot M_i(q_i)-2C_i(q_i,\dot q_i)$ (see, e.g., \cite{Slotine1991_Book,Spong2006_Book}) being utilized, $\forall i$.
This leads us to immediately obtain that $s_i\in{\mathcal L}_2\cap {\mathcal L}_\infty$ and $\hat\vartheta_i\in {\mathcal L}_\infty$, $\forall i$.
By introducing the following sliding vector (the same as \cite{Chopra2006})
\be
\label{eq:47}
\xi_i=\dot q_i+\alpha q_i,
\ee
we can rewrite the first subsystem of (\ref{eq:46}) as
\be
\label{eq:48}
\dot \xi_i=-\Sigma_{j\in{\mathcal N}_i(t)}w_{ij}(t)(\xi_i-\xi_j)+\lambda_{\mathcal M} s_i+\dot s_i.
\ee
The combination of all the equations like (\ref{eq:48}) gives
\be
\label{eq:49}
\dot \xi=-[{\mathcal L}_w(t)\otimes I_m]\xi+\lambda_{\mathcal M} s^\ast+\dot s^\ast
\ee
where $\xi=[\xi_1^T,\dots,\xi_n^T]^T$ and $s^\ast=[s_1^T,\dots,s_n^T]^T$, $\otimes$ denotes the Kronecker product \cite{Brewer1978_TCS}, and $I_m$ denotes the $m\times m$ identity matrix. By following the standard practice (see, e.g., \cite{Lee2007_TAC,Ren2008_Book}), we introduce two vectors
$\xi_E=[\xi_1^T-\xi_2^T,\dots,\xi_{n-1}^T-\xi_n^T]^T$ and $s_E=[s_1^T-s_2^T, \dots,s_{n-1}^T -s_n^T]^T$, upon which we obtain from (\ref{eq:49}) that (by following \cite{Ren2008_Book})
\be
\label{eq:50}
\dot \xi_E=-\Omega(t)\xi_E+\lambda_{\mathcal M} s_E+\dot s_E
\ee
with $s_E\in{\mathcal L}_2\cap{\mathcal L}_\infty$ and $\Omega(t)$ being a time-varying matrix that is determined by $\mathcal L_w(t)$. In accordance with \cite[p.~48,~p.~49]{Ren2008_Book}, the system (\ref{eq:50}) with $s_E=0$ and $\dot s_E=0$ 
is uniformly exponentially stable. By the standard linear system theory, the output $\xi_E$ of (\ref{eq:50}) can be considered as the superposition of the output $\xi_E^\ast$ of
\be
\label{eq:52}
\dot \xi_E^{\ast}=-\Omega(t)\xi_E^{\ast}+\lambda_{\mathcal M}s_E
\ee
and the output $\xi_E^{\ast\ast}$ of
\be
\label{eq:53}
\dot \xi_E^{\ast\ast}=-\Omega(t)\xi_E^{\ast\ast}+\dot s_E.
\ee
For the system (\ref{eq:52}) with $\lambda_{\mathcal M}s_E$ as the input, we obtain from the standard input-output properties of uniformly exponentially stable linear systems  that $\xi_E^\ast\in{\mathcal L}_2\cap{\mathcal L}_\infty$ since $\lambda_{\mathcal M} s_E\in{\mathcal L}_2\cap{\mathcal L}_\infty$. For the system (\ref{eq:53}) with $\dot s_E$ as the input, we obtain from Lemma 2  that $\xi_E^{\ast\ast}\in{\mathcal L}_2\cap{\mathcal L}_\infty$ since $\int_0^t \dot s_E(\sigma)d\sigma+s_E(0)=s_E\in{\mathcal L}_2$ and $\int_0^t \dot s_E(\sigma)d\sigma=s_E-s_E(0)\in{\mathcal L}_\infty$. Hence $\xi_E\in{\mathcal L}_2\cap{\mathcal L}_\infty$. Using (\ref{eq:43}), we can rewrite (\ref{eq:42}) as
\be
\label{eq:54}
\dot z_i=-\alpha z_i+\underbrace{(\lambda_{\mathcal M}-\alpha)s_i-\Sigma_{j\in{\mathcal N}_i(t)}w_{ij}(t)(\xi_i-\xi_j)}_{\psi_i^\ast}
\ee
where $\psi _i^\ast\in{\mathcal L}_2\cap{\mathcal L}_\infty$, $\forall i$. Using the input-output properties of exponentially stable and strictly proper linear systems \cite[p.~59]{Desoer1975_Book}, we obtain from (\ref{eq:54}) that $z_i\in{\mathcal L}_2\cap{\mathcal L}_\infty$, $\dot z_i\in{\mathcal L}_2\cap{\mathcal L}_\infty$, and $z_i\to 0$ as $t\to\infty$, $\forall i$. This leads us to directly obtain from (\ref{eq:43}) that $\dot q_i\in{\mathcal L}_2\cap \mathcal L_\infty$, $\forall i$. From the second subsystem of (\ref{eq:46}) and using the property that $M_i(q_i)$ is uniformly positive definite (see, e.g., \cite{Slotine1991_Book,Spong2006_Book}), we obtain that $\dot s_i\in{\mathcal L}_\infty$, and as a consequence, $\ddot q_i\in{\mathcal L}_\infty$, $\forall i$. Using (\ref{eq:47}), we can directly express $\xi_E$ as
\be
\label{eq:55}
\xi_E=\dot q_E+\alpha q_E
\ee
with $q_E=[q_1^T-q_2^T,\dots,q_{n-1}^T-q_n^T]^T$, and equation (\ref{eq:55}) can further be written as
\be
\label{eq:56}
\dot q_E=-\alpha q_E+\xi_E.
\ee
For the system (\ref{eq:56}) with $\xi_E$ as the input and $q_E$ as the output, using the input-output properties of exponentially stable and strictly proper linear systems \cite[p.~59]{Desoer1975_Book} yields the result that $q_E\in{\mathcal L}_2\cap{\mathcal L}_\infty$, $\dot q_E\in{\mathcal L}_2\cap{\mathcal L}_\infty$, and $q_E\to 0$ as $t\to\infty$, which immediately gives the result that $q_i-q_j\to 0$ as $t\to\infty$, $\forall i,j$. The result that $\ddot q_i\in{\mathcal L}_\infty$ implies that $\dot q_i$ is uniformly continuous, $\forall i$. Using the properties of square-integrable and uniformly continuous functions \cite[p.~232]{Desoer1975_Book}, we obtain that $\dot q_i\to 0$ as $t\to\infty$, $\forall i$.

We next demonstrate that the manipulability of the system is infinite with degree one if an external physical input action is exerted at the torque level on a system that acts as the root of the interaction graph (in the sense that there exist an infinite number of uniformly bounded intervals such that the system acts as the root of the union of the interaction graphs in each interval). Without loss of generality, suppose that the $\kappa$-th system acts as the root of the interaction graph and is subjected to an external physical input action $\tau_{h,\kappa}$, and we then have that
\be
\begin{cases}
M_{\kappa}(q_{\kappa})\dot s_\kappa+C_\kappa(q_\kappa,\dot q_\kappa)s_\kappa\\
=-K_\kappa s_\kappa+Y_\kappa(q_\kappa,\dot q_\kappa,z_\kappa,\dot z_\kappa)\Delta \vartheta_\kappa+\tau_{h,\kappa}\\
\dot{\hat \vartheta}_\kappa=-\Gamma_\kappa Y_\kappa^T(q_\kappa,\dot q_\kappa,z_\kappa,\dot z_\kappa)s_\kappa.
\end{cases}
\ee
The derivative of $V_\kappa$ along the trajectories of the system now becomes (by using the standard basic inequalities)
\be
\dot V_\kappa=-s_\kappa ^TK_\kappa s_\kappa+s_\kappa^T\tau_{h,\kappa}\le -\frac{1}{2}s_\kappa^T K_\kappa s_\kappa+\frac{1}{2}\tau_{h,\kappa}^T K_{\kappa}^{-1}\tau_{h,\kappa}
\ee
and this implies that the $\mathcal L_2$-gain from $\tau_{h,\kappa}$ to $s_\kappa$ is less than or equal to ${1}/{\lambda_{\min}\{K_\kappa\}}$. The $\mathcal L_2$-gain from $s^\ast$ to $\lambda_{\mathcal M}s^\ast+\dot s^\ast$, similar as before, can be directly obtained by calculating the $\mathcal H_\infty$ norm of the transfer function as $\sqrt{\lambda_{\mathcal M}^2+(\sup_\omega|\omega|)^2}$. For the system (\ref{eq:49}) with $s^\ast=0$ and $\dot s^\ast=0$, $\xi$ uniformly asymptotically converges to certain constant vector in accordance with \cite{Ren2005_TAC,Ren2008_Book}, and thus the system (\ref{eq:49}) with $s^\ast=0$ and $\dot s^\ast=0$ is a uniformly marginally stable linear system of the first kind (i.e., with the state uniformly converging to a constant vector) by the standard linear system theory. It can then be directly shown from Lemma 3 that the $\mathcal L_2$-gain from $\lambda_{\mathcal M}s^\ast+\dot s^\ast$ to $\dot \xi$ is finite, and by letting $\xi_c=(1/n)\Sigma_{i=1}^n \xi_i
$, we directly obtain that the $\mathcal L_2$-gain from $\lambda_{\mathcal M}s^\ast+\dot s^\ast$ to $\dot \xi_c$ is less than or equal to a positive constant $h^\ast$ (i.e., finite). The $\mathcal L_2$-gain from $\dot \xi_c$ to $ \xi_c-\xi_c(0)$ is $\sup_\omega\frac{1}{|\omega|}$. Then we obtain the $\mathcal L_2$-gain from $\tau_{h,\kappa}$ to $\xi_c-\xi_c(0)$ as
\begin{align}
{\mathcal M}_{\tau_{h,\kappa}\mapsto \xi_c-\xi_c(0)}\le&\frac{h^\ast}{\lambda_{\min}\{K_\kappa\}}\sqrt{\lambda_{\mathcal M}^2+(\sup_\omega|\omega|)^2}\sup_\omega\frac{1}{|\omega|}\nn\\
=&\frac{h^\ast}{\lambda_{\min}\{K_\kappa\}}\sqrt{\lambda_{\mathcal M}^2\sup_\omega\frac{1}{|\omega|^2}+1}=\infty.
\end{align}
 Let $q_c=(1/n)\Sigma_{k=1}^n q_k$, and by exploiting the relation that $\xi_c=\dot q_c+\alpha q_c$ [obtained from (\ref{eq:47})] with the $\mathcal L_2$-gain from $\xi_c-\xi_c(0)$ to $q_c-[q_c(0)+(1/\alpha)\dot q_c(0)]$ being $1/\alpha$ (which is obtained, similar as before, by calculating the $\mathcal H_\infty$ norm of the transfer function), the $\mathcal L_2$-gain from $\tau_{h,\kappa}$ to $q_c-[q_c(0)+(1/\alpha)\dot q_c(0)]$ thus satisfies \begin{align}
&{\mathcal M}_{\tau_{h,\kappa}\mapsto q_c-[q_c(0)+(1/\alpha)\dot q_c(0)]}\nn\\
&\le \frac{h^\ast}{\alpha \lambda_{\min}\{K_\kappa\}}\sqrt{\lambda_{\mathcal M}^2\sup_\omega\frac{1}{|\omega|^2}+1}=\infty.
\end{align}
This implies that the infinite manipulability of the system with degree one is guaranteed. \hfill{\small {$\blacksquare $}}

\subsection{Consensus With Switching Topology and Communication Delays}

For addressing the more complicated case that both the switching directed topologies and unknown time-varying communication delays are involved, we define the vector $z_i$ by
\begin{align}
\label{eq:61}
\dot z_i=&-\alpha \dot q_i+\lambda_{\mathcal M} (\dot q_i-z_i)\nn\\
&-\Sigma_{j\in\mathcal N_i(t)} w_{ij}(t)[\xi_i-\xi_j(t-T_{ij})]
\end{align}
where $T_{ij}$ is the time-varying communication delay from the $j$-th system to the $i$-th system. The communication delay is assumed to be rather general in the sense that it is piecewise uniformly continuous and uniformly bounded. The adaptive controller remains the same as the one given by (\ref{eq:44}) and (\ref{eq:45}) yet with $z_i$ being given by (\ref{eq:61}).

\emph{Theorem 4:} Suppose that there exist an infinite number of uniformly bounded intervals $[t_{i_\ell},t_{i_{\ell+1}})$, $\ell=1,2,\dots$ with $t_{i_1}=t_0$ satisfying the property that the union of the interaction graphs in each interval contains a directed spanning tree and that the time-varying communication delays are piecewise uniformly continuous and uniformly bounded. Then the adaptive controller given by (\ref{eq:44}) and (\ref{eq:45}) with $z_i$ being given by (\ref{eq:61}) ensures 1) the consensus of the $n$ systems in free motion (i.e., no external physical interaction), i.e., $q_i-q_j\to 0$ and $\dot q_i\to 0$ as $t\to \infty$, $\forall i,j=1,\dots,n$ and 2) the infinite manipulability of the system with degree one in terms of an external physical input action at the torque level and the consensus equilibrium increment.

\emph{Proof:} The proof of Theorem 4 is quite similar to that of Theorem 3 except that the interconnected system (\ref{eq:48}) now becomes
\be
\label{eq:62}
\dot \xi_i=-\Sigma_{j\in{\mathcal N}_i(t)}w_{ij}(t)[\xi_i-\xi_j(t-T_{ij})]+\lambda_{\mathcal M} s_i+\dot s_i
\ee
where the time-varying communication delays are involved, and in addition, $s_i\in{\mathcal L}_2\cap{\mathcal L}_\infty$ and $\hat \vartheta_i\in{\mathcal L}_\infty$ (which is guaranteed by the adaptive controller, similar to the proof of Theorem 3), $\forall i$. All the equations expressed as (\ref{eq:62}) can be written compactly as
\be
\label{eq:63}
\dot \xi={\mathcal F}_D(\xi)+\lambda_{\mathcal M}s^\ast+\dot s^\ast
\ee
with ${\mathcal F}_D(\cdot)$ denoting a linear mapping that involves delay operation, and the output is specified to be the same as the one in (\ref{eq:50}), which can be explicitly expressed as (see, e.g., \cite{Lee2007_TAC})
\be
\label{eq:64}
\xi_E=(\bar C\otimes I_m)\xi
\ee
where $\bar C\in R^{(n-1)\times n}$ is given as
\be
\label{eq:65}
\bar C=\begin{bmatrix}1 & -1 & 0 &\dots & 0\\
0 & 1 & -1 &\dots & 0\\
\vdots & \ddots & \ddots &\ddots & \vdots\\
0& \dots & 0 & 1 & -1\end{bmatrix}.
\ee
In accordance with the result in \cite{Munz2011b_TAC}, the output $\xi_E$ of the linear system given by (\ref{eq:63}) and (\ref{eq:64}) with $s^\ast=0$ and $\dot s^\ast=0$ uniformly asymptotically converges to zero (with $\xi$ uniformly asymptotically converging to a constant vector) and furthermore $\dot \xi $ also uniformly asymptotically converges to zero (in the case that $s^\ast=0$ and $\dot s^\ast=0$). Therefore, by the standard linear system theory, the system (\ref{eq:63}) with $s^\ast=0$ and $\dot s^\ast=0$ is a uniformly marginally stable linear system of the first kind (i.e., with the state uniformly converging to a constant vector), and in addition, it is well known from the standard linear system theory that both $\xi_E$ and $\dot \xi$ uniformly exponentially converge to zero in the case that $s^\ast=0$ and $\dot s^\ast=0$. For the system given by (\ref{eq:63}) and (\ref{eq:64}), using Lemma 1, the standard input-output properties of linear time-varying systems, and the standard superposition principle (with $\lambda_{\mathcal M}s^\ast$ and $\dot s^\ast$, respectively, as the input), we obtain that $\xi_E\in{\mathcal L}_\infty$. For the system given by (\ref{eq:63}), using Lemma 3 and the standard superposition principle (with $\lambda_{\mathcal M}s^\ast$ and $\dot s^\ast$, respectively, as the input), we obtain that ${\mathcal F}_D(\xi)\in{\mathcal L}_\infty$. Equation (\ref{eq:61}) can be rewritten as (using the definition of $s_i$)
\begin{align}
\dot z_i=&-\alpha z_i+(\lambda_{\mathcal M}-\alpha) s_i\nn\\
&-\Sigma_{j\in\mathcal N_i(t)} w_{ij}(t)[\xi_i-\xi_j(t-T_{ij})]
\end{align}
which yields the result that $z_i\in{\mathcal L}_\infty$ and $\dot z_i\in{\mathcal L}_\infty$ in accordance with the input-output properties of exponentially stable and strictly proper linear systems \cite[p.~59]{Desoer1975_Book}, $\forall i$. Thus, $\dot q_i\in{\mathcal L}_\infty$, $\forall i$. We then obtain that $\dot s_i\in{\mathcal L}_\infty$ from the following equation [i.e., the second subsystem of (\ref{eq:46}) with $z_i$ and $\dot z_i$ being given by (\ref{eq:61})]
\be
\label{eq:aa3}
M_i(q_i)\dot s_i+C_i(q_i,\dot q_i)s_i=-K_i s_i+Y_i(q_i,\dot q_i,z_i,\dot z_i)\Delta \vartheta_i
\ee
  and by using the property that $M_i(q_i)$ is uniformly positive definite (see, e.g., \cite{Slotine1991_Book,Spong2006_Book}), $\forall i$. This implies that $\ddot q_i\in{\mathcal L}_\infty$, $\forall i$. Hence, $\dot q_i$ and $s_i$ are uniformly continuous, $\forall i$. Using the properties of square-integrable and uniformly continuous functions \cite[p.~232]{Desoer1975_Book}, we obtain that $s_i\to 0$ as $t\to\infty$, $\forall i$. The result that $\dot q_i\in{\mathcal L}_\infty$, $\dot z_i\in{\mathcal L}_\infty$, and $\dot{\hat\vartheta}_i\in{\mathcal L}_\infty$ [from (\ref{eq:45}) with $z_i$ and $\dot z_i$ being given by (\ref{eq:61})] implies that $q_i$, $z_i$, and $\hat\vartheta_i$ are all uniformly continuous, $\forall i$. Then we obtain from (\ref{eq:61}) that $\dot z_i$ is piecewise uniformly continuous by additionally considering the assumption that the time-varying delays are piecewise uniformly continuous and uniformly bounded and the standard assumption concerning the dwell time (i.e., $t_{k+1}-t_k\ge T_D$, $\forall k=0,1,\dots$), $\forall i$. From (\ref{eq:aa3}) and using the aforementioned property that $M_i(q_i)$ is uniformly positive definite, we obtain that $\dot s_i$ is piecewise uniformly continuous, $\forall i$. The application of the standard generalized Barbalat's Lemma (see, e.g., \cite{Jiang2009_IET}) immediately yields the result that $\dot s_i\to 0$ as $t\to\infty$, $\forall i$. For the system given by (\ref{eq:63}) and (\ref{eq:64}) with $\lambda_{\mathcal M}s^\ast+\dot s^\ast$ as the input, we obtain from the standard input-output properties of linear time-varying systems that $\xi_E\to 0$ as $t\to\infty$, and we also obtain from Lemma 3 that $\dot \xi\to 0$ as $t\to\infty$. From the definition of $\xi_i$ given by (\ref{eq:47}), we directly obtain that $\ddot q_i=-\alpha\dot q_i+\dot \xi_i$, upon which, we obtain from the input-output properties of exponentially stable and strictly property linear systems \cite[p.~59]{Desoer1975_Book} that $\dot q_i\to 0$ as $t\to\infty$, $\forall i$. From (\ref{eq:56}) and using the input-output properties of exponentially stable and strictly property linear systems \cite[p.~59]{Desoer1975_Book}, we obtain that $q_E\to 0$ as $t\to\infty$, and therefore, $q_i-q_j\to 0$ as $t\to\infty$, $\forall i,j$.


 The proof of the second part of Theorem 4 can be performed by following similar procedures as in that of Theorem 3. \hfill{\small {$\blacksquare $}}

\emph{Remark 4:} The definition of $z_i$ by the dynamic system (\ref{eq:42}) or (\ref{eq:61}) is motivated by
but different from \cite{Wang2017_CAC,Wang2017_AUTSubmitted} in the sense that $z_i$ is no longer the pure integration concerning the
system state. This is reflected in the newly introduced term
$\lambda_{\mathcal M}(\dot q_i-z_i)=\lambda_{\mathcal M}s_i$, which, as is shown,
is crucial for guaranteeing the high (infinite) manipulability
of the system. 

\emph{Remark 5:} The proposed adaptive controllers can also ensure the position consensus of the networked Lagrangian systems or the asymptotic convergence of the velocity of a single Lagrangian system (i.e., the case of manipulation of a single
Lagrangian system in Sec. IV) in the case that the external input action is square-integrable, as can directly be observed from the previous analysis. This, in turn, fundamentally ensures that the external subject (e.g., a human operator) can easily manipulate the interactive systems and simultaneously that the asymptotic consensus among the systems or convergence of the velocity of the system be maintained. An intuitive interpretation concerning the possibility of simultaneously achieving the two objectives is tightly associated with the properties of functions that are ``square-integrable yet not integral bounded''. A well-known function that is square-integrable yet not integral bounded is $$f(t)=\frac{1}{t+1},$$
   and its integral can be directly demonstrated to satisfy the well-recognized property that $$\int_0^t f(\sigma)d\sigma=\ln(t+1)\to \infty$$ as $t\to\infty$. In particular, due to the square-integrability of the external input action, the asymptotic consensus among the systems or convergence of the velocity of the system is maintained even under the external input action, and due to the possibility of integral unboundedness of the external input action, manipulating the system to an arbitrary equilibrium without using so much effort (i.e., with finite energy consumption) becomes possible. 

\section{Application to Bilateral Teleoperation With Time-Varying Delay}

Bilateral teleoperation with arbitrary unknown time-varying communication delay is a longstanding benchmark problem in the literature, and to the best of our knowledge, no delay-independent solutions have been reported and systematically developed. The standard scattering/wave-variable-based approach \cite{Anderson1989_TAC,Niemeyer1991_JOE} can typically handle arbitrary unknown constant communication delay. The modifications to the original scattering/wave-variable-based approach appear in, e.g., \cite{Niemeyer1998_ICRA,Yokokohji1999_IROS,Chopra2003_ACC,Munir2002_TMECH,Ching2006_JDSMC,Chopra2008_TCST} for handling time-varying delay or position drift. It is typically recognized that the scattering/wave-variable-based approach exhibits potential limitations as handling the problem of position drift (see, e.g., \cite{Lee2006_TRO,Chopra2006_TRO}). For resolving this problem, numerous synchronization-based results under constant or time-varying delay are presented (see, for instance, \cite{Chopra2006_TRO,Lee2006_TRO,Chopra2008_AUT,Nuno2009_IJRR,
Polushin2006_TCybernetics,Lee2010_TRO_PassiveSetPosition,Nuno2010_AUT,Liu2015_TMECH,Polushin2015_TMECH,Nuno2018_IJRNC,Hua2017_TAC}). 
However, most of these results, as handling the case that the delay is time-varying, are generally delay-dependent.

 Here we provide a delay-independent (i.e., independent of the arbitrary time-varying delay in the sense that the delay is only required to be piecewise uniformly continuous and uniformly bounded) solution to this longstanding open problem and this solution can be considered as a special case of the result in Sec. V. The distinguished point of the proposed solution here, in contrast with those in the literature, is the
appropriate composite of using a new class of dynamic feedback and guaranteeing the infinite manipulability of the teleoperator system. Specifically, we consider a teleoperator system consisting of two robots with their dynamics being given by \cite{Hokayem2006_AUT,Spong2006_Book}
\begin{align}
\label{eq:74}
M_1(q_1)\ddot q_1+C_1(q_1,\dot q_1)\dot q_1+g_1(q_1)=\tau_1+\tau^\ast_1\\
\label{eq:75}
M_2(q_2)\ddot q_2+C_2(q_2,\dot q_2)\dot q_2+g_2(q_2)=\tau_2-\tau^\ast_2
\end{align}
with $\tau^\ast_1$ the torque exerted by the human operator on the master robot (the 1st robot) and $\tau^\ast_2$ the torque exerted by the slave robot (the 2nd robot) on the environment. The adaptive controller is given as
\begin{align}
\label{eq:76}
\tau_1=&-Ks_1+Y_1(q_1,\dot q_1,z_1,\dot z_1)\hat \vartheta_1\\
\label{eq:77}
\dot{\hat \vartheta}_1=&-\Gamma_1 Y_1^T(q_1,\dot q_1,z_1,\dot z_1) s_1\\
\label{eq:78}
\tau_2=&-Ks_2+Y_2(q_2,\dot q_2,z_2,\dot z_2)\hat \vartheta_2\\
\label{eq:79}
\dot{\hat \vartheta}_2=&-\Gamma_2 Y_2^T(q_2,\dot q_2,z_2,\dot z_2) s_2
\end{align}
with $z_1$ and $z_2$ being defined as
\begin{align}
\label{eq:80}
\dot z_1=&-\alpha \dot q_1+\lambda_{\mathcal M} (\dot q_1-z_1)-\lambda[\xi_1-\xi_2(t-T_{12})]\\
\label{eq:81}
\dot z_2=&-\alpha \dot q_2+\lambda_{\mathcal M} (\dot q_2-z_2)-\lambda[\xi_2-\xi_1(t-T_{21})]
\end{align}
where $K$ is a symmetric positive definite matrix and $\lambda$ is a positive design constant.

\emph{Theorem 5:} Suppose that the time-varying communication delays are piecewise uniformly continuous and uniformly bounded. Then the adaptive controller given by (\ref{eq:76}), (\ref{eq:77}), (\ref{eq:78}), and (\ref{eq:79}) with $z_1$ and $z_2$ being respectively given by (\ref{eq:80}) and (\ref{eq:81}) for the teleoperator system given by (\ref{eq:74}) and (\ref{eq:75}) ensures position synchronization of the master and slave robots in free motion (i.e., $\tau_1^\ast=\tau_2^\ast=0$) and static torque reflection without considering the gravitational torque estimation errors. In addition, the manipulability of the teleoperator system is infinite with degree one.

 The proof of Theorem 5 can be performed by following similar procedures as in those of Theorem 3 and Theorem 4. The special issue that needs to be further demonstrated in the case of bilateral teleoperation is that of force/torque reflection, and the analysis of the torque reflection property of the teleoperator system can be completed by following the standard practice (see, e.g., \cite{Lee2006_TRO,Liu2013_AUT,Liu2015_TMECH}). In particular, consider the scenario that $\dot q_i$, $\ddot q_i$, and $\dot z_i$ converge to zero, $i=1,2$, in which case it can be directly shown that
 \begin{align}
 \tau_1^\ast=&\frac{\lambda\alpha K}{\lambda_{\mathcal M}} (q_1-q_2)-[\hat g_1(q_1)-g_1(q_1)]\\
 \tau_2^\ast=&-\frac{\lambda\alpha K}{\lambda_{\mathcal M}} (q_2-q_1)+[\hat g_2(q_2)-g_2(q_2)]
 \end{align}
 with $\hat g_i(q_i)$ being the estimate of $g_i(q_i)$, $i=1,2$. This then leads us to straightforwardly obtain that if there are no gravitational torque estimation errors, then $\tau_1^\ast=\tau_2^\ast$, i.e., the static torque reflection in the sense of \cite{Lee2006_TRO} is achieved.

\emph{Remark 6:} Our result is in contrast to those in \cite{Abdessameud2014_TAC} and \cite{Nuno2009_IJRR} which rely on the conditions associated with the choice of the gains based on some a priori information of the time-varying delay, and in particular, neither the upper bound of the delay nor that of the discontinuous change of the delay is required. The standard assumption concerning the derivative of the time-varying delay in, e.g.,  \cite{Niemeyer1998_ICRA,Chopra2003_ACC} is also no longer required. 

\section{Simulation Results}

\subsection{Teaching Operation of A Single Robotic System}

Consider a standard two-DOF planar robot with the adaptive controller given in Sec. IV-B being exerted.  The controller parameters are chosen as $K=16.0I_2$, $\alpha=2.0$, $\Gamma=8.0I_3$, and $\lambda_{\mathcal M}=2.0$. The initial parameter estimate is set as $\hat\vartheta(0)=[0,0,0]^T$. Suppose that the controlled robotic system is subjected to the manipulation action by a human operator with the exerted control torque being modeled as the standard PD control, i.e., $\tau_h=-5.0 \dot q-10.0(q-q_h)$ with $q_h=[3.5,3.0]^T$ being the desired position. The sampling period is set as 5 ms. The simulation results are shown in Fig. 1 and Fig. 2. As is shown, the position of the robot is manipulated to the desired one and in addition, the exerted control torques of the human operator asymptotically converge to zero (implying that the human operator does not need to constantly maintain a holding torque). For comparison, we also conduct a simulation with $\lambda_{\mathcal M}=0$ (in which case, the manipulability of the system is finite rather than infinite), and as shown in Fig. 3, it becomes difficult for the human operator to manipulate the robot to the desired position.

\begin{figure}
\centering
\begin{minipage}[t]{1.0\linewidth}
\centering
\includegraphics[width=3.0in]{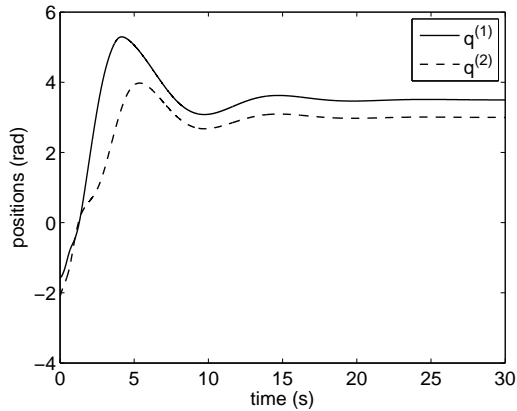}
\caption{Positions of the two-DOF robot with $\lambda_{\mathcal M}=6.0$.}
\end{minipage}%
\end{figure}

\begin{figure}
\centering
\begin{minipage}[t]{1.0\linewidth}
\centering
\includegraphics[width=3.0in]{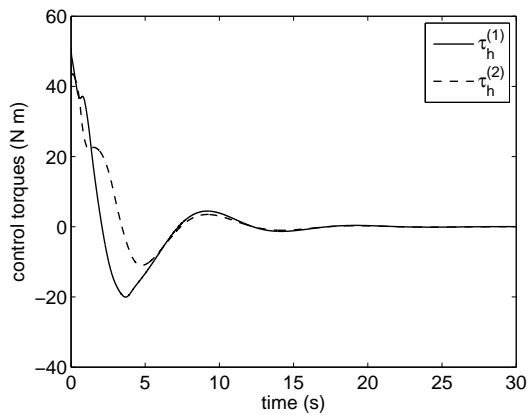}
\caption{Control torques exerted by the human operator with $\lambda_{\mathcal M}=6.0$.}
\end{minipage}%
\end{figure}

\begin{figure}
\centering
\begin{minipage}[t]{1.0\linewidth}
\centering
\includegraphics[width=3.0in]{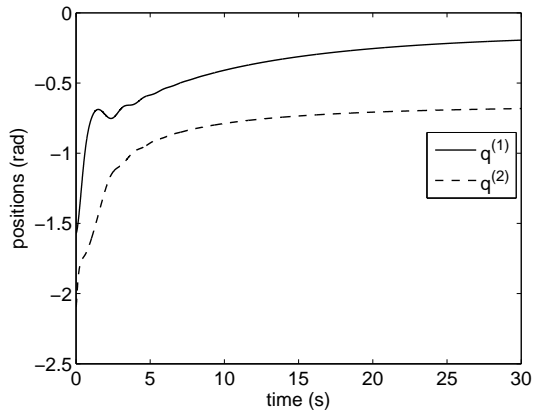}
\caption{Positions of the two-DOF robot with $\lambda_{\mathcal M}=0$.}
\end{minipage}%
\end{figure}

\subsection{Networked Robotic Systems}

Let us consider six standard two-DOF planar robots and the interaction graph among the six robots randomly switches among the ones given in Fig. 4. The interaction graph randomly switches among the three graphs in Fig. 4 every 150 ms, in accordance with the uniform distribution and the union of the three graphs is shown in Fig. 5.

 The initial joint positions of the robots are set as $q_1(0)=[-\pi/3,-\pi/2]^T$, $q_2(0)=[-2\pi/3,\pi/3]^T$, $q_3(0)=[5\pi/6,-\pi/3]^T$, $q_4(0)=[\pi/6,\pi/2]^T$, $q_5(0)=[\pi/2,\pi/6]^T$, and $q_6(0)=[-\pi/6,-\pi/3]^T$. The initial joint velocities of the robots are set as $\dot q_i(0)=[0,0]^T$, $i=1,\dots,6$. The controller parameters are chosen as $K_i=16.0 I_2$, $\alpha=1.6$, $\Gamma_i=8.0 I_3$, $i=1,\dots,6$, and $\lambda_{\mathcal M}=10.0$. The initial value of $z_i$ is set as $z_i(0)=[0,0]^T$, $i=1,\dots,6$. The adjacency weights are set as $w_{ij}(t)=1.0$ if $j\in{\mathcal N}_i(t)$, and $w_{ij}(t)=0$ otherwise, $\forall i,j=1,\dots,6$. The initial parameter estimates are set as $\hat \vartheta_i(0)=[0,0,0]^T$, $i=1,\dots,6$. Suppose that the 3rd robot is manipulated by a human operator with the standard PD control action $\tau_{h,3}=-5.0\dot q_3-10.0(q_3-q_h)$ with $q_h=[3.5,3.0]^T$ being the desired position. The simulation results are shown in Fig. 6 and Fig. 7 and the positions of the robots apparently converge to $q_h$, which implies that the human operator does not need to hold the 3rd robot with a torque constantly, due to the infinite manipulability of the closed-loop system. In fact, $\tau_{h,3}$ converges to zero as the positions of the robots converge to $q_h$. To highlight the role of the action $\lambda_{\mathcal M}s_i$ employed in the controller, we perform another simulation under the same context except that $\lambda_{\mathcal M}$ is set as $\lambda_{\mathcal M}=0$. The simulation results are shown in Fig. 8 and Fig. 9, from which we observe that the positions of the robots do not converge to $q_h$. This in turn implies that the holding torque of the human operator does not converge to zero.

 We next perform a simulation with the time-varying communication delay being taken into account. The controller parameters are chosen to be same as above except that $\Gamma_i$ is reduced to $\Gamma_i=1.0 I_3$, $i=1,\dots,6$. The time-varying communication delays are set as $T_{ij}(t)=0.3+\eta_{ij}$ with $\eta_{ij}$ conforming to the uniform distribution over the interval $[0,0.9]$ and changes every 30 ms, $j\in{\mathcal N}_i(t)$, $i=1,\dots,6$. The simulation results are shown in Fig. 10 and Fig. 11, and we can observe that the positions of the robots indeed converge to the desired one ($q_h$).

\begin{figure}
\centering
\begin{minipage}[t]{1.0\linewidth}
\centering
\includegraphics[width=3.0in]{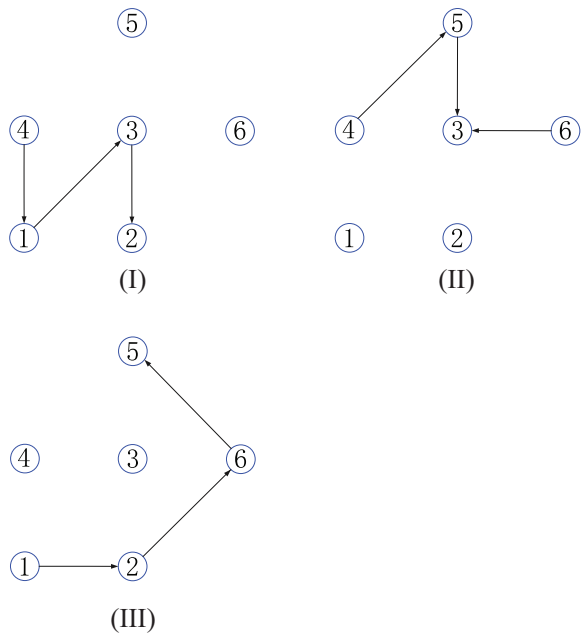}
\caption{Three interaction graphs among the six robots.}
\end{minipage}%
\end{figure}

\begin{figure}
\centering
\begin{minipage}[t]{1.0\linewidth}
\centering
\includegraphics[width=2.0in]{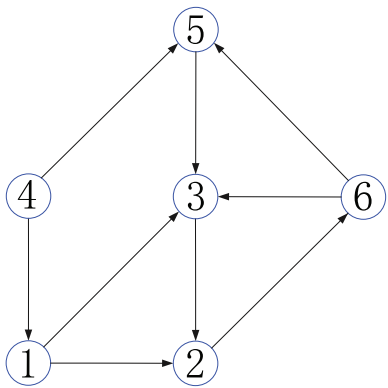}
\caption{Union of the interaction graphs.}
\end{minipage}%
\end{figure}

\begin{figure}
\centering
\begin{minipage}[t]{1.0\linewidth}
\centering
\includegraphics[width=3.2in]{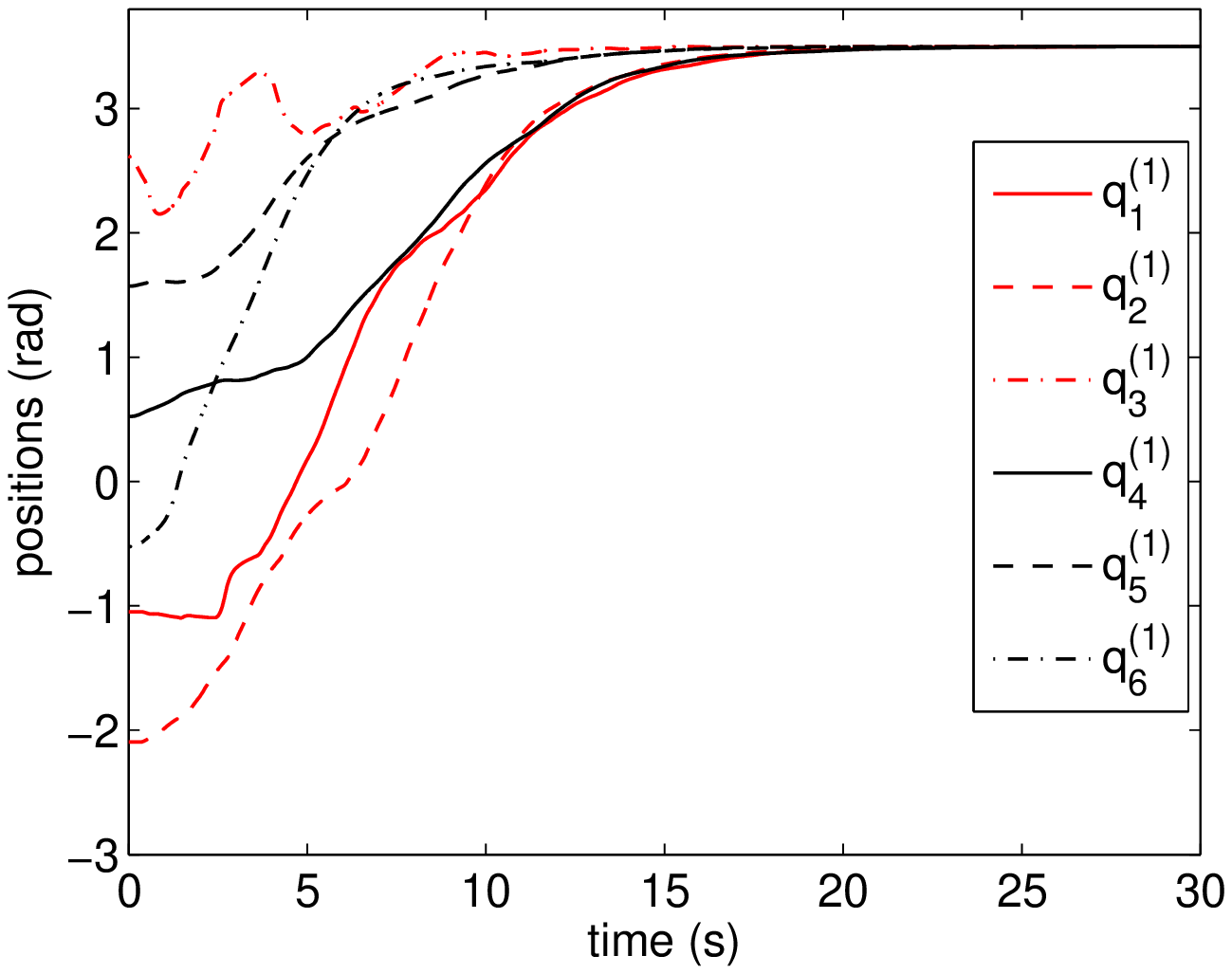}
\caption{Positions of the robots with the 3rd robot being manipulated by a human operator with $\lambda_{\mathcal M}=10.0$ (first coordinate).}
\end{minipage}%
\end{figure}

\begin{figure}
\centering
\begin{minipage}[t]{1.0\linewidth}
\centering
\includegraphics[width=3.2in]{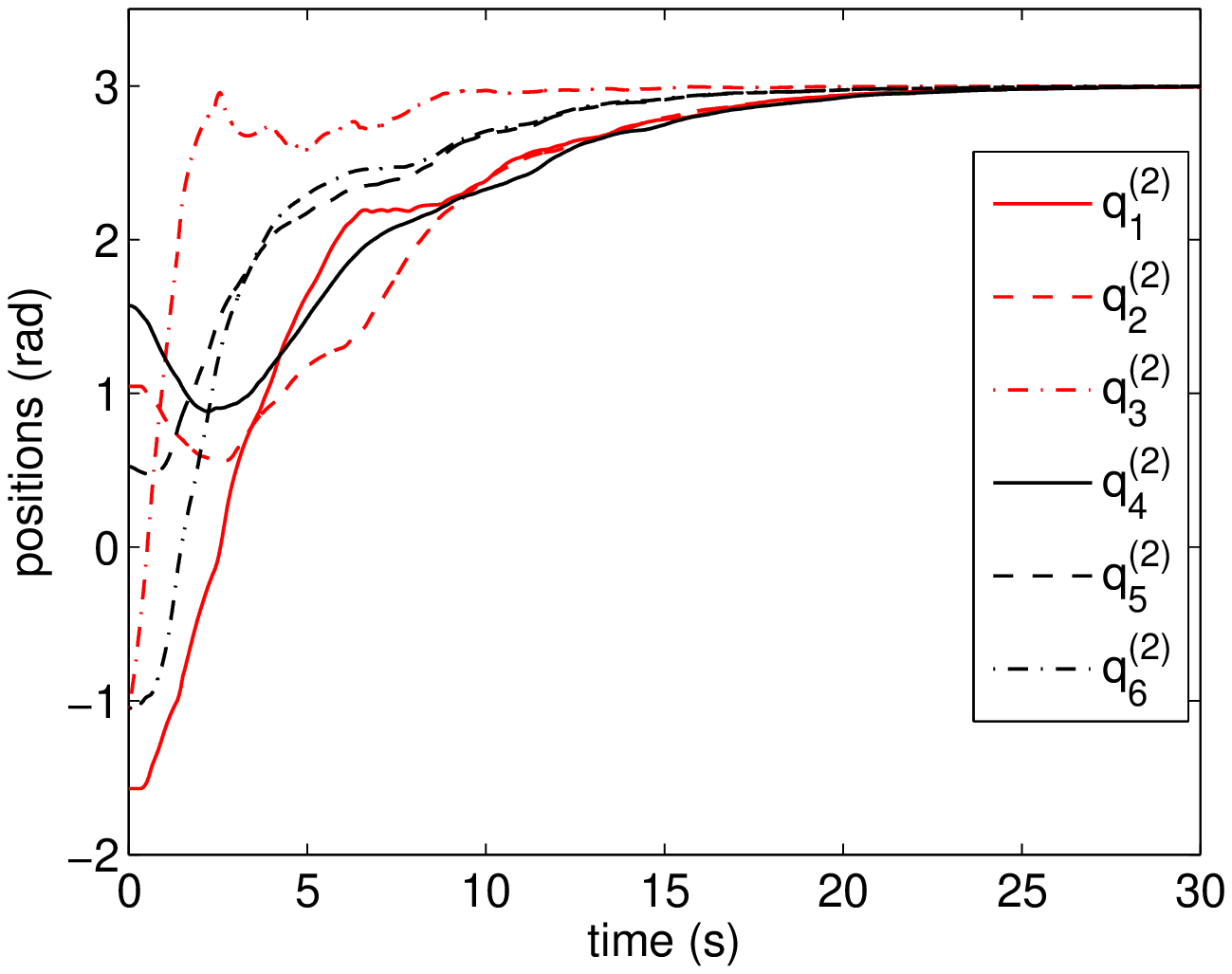}
\caption{Positions of the robots with the 3rd robot being manipulated by a human operator $\lambda_{\mathcal M}=10.0$ (second coordinate).}
\end{minipage}%
\end{figure}

\begin{figure}
\centering
\begin{minipage}[t]{1.0\linewidth}
\centering
\includegraphics[width=3.2in]{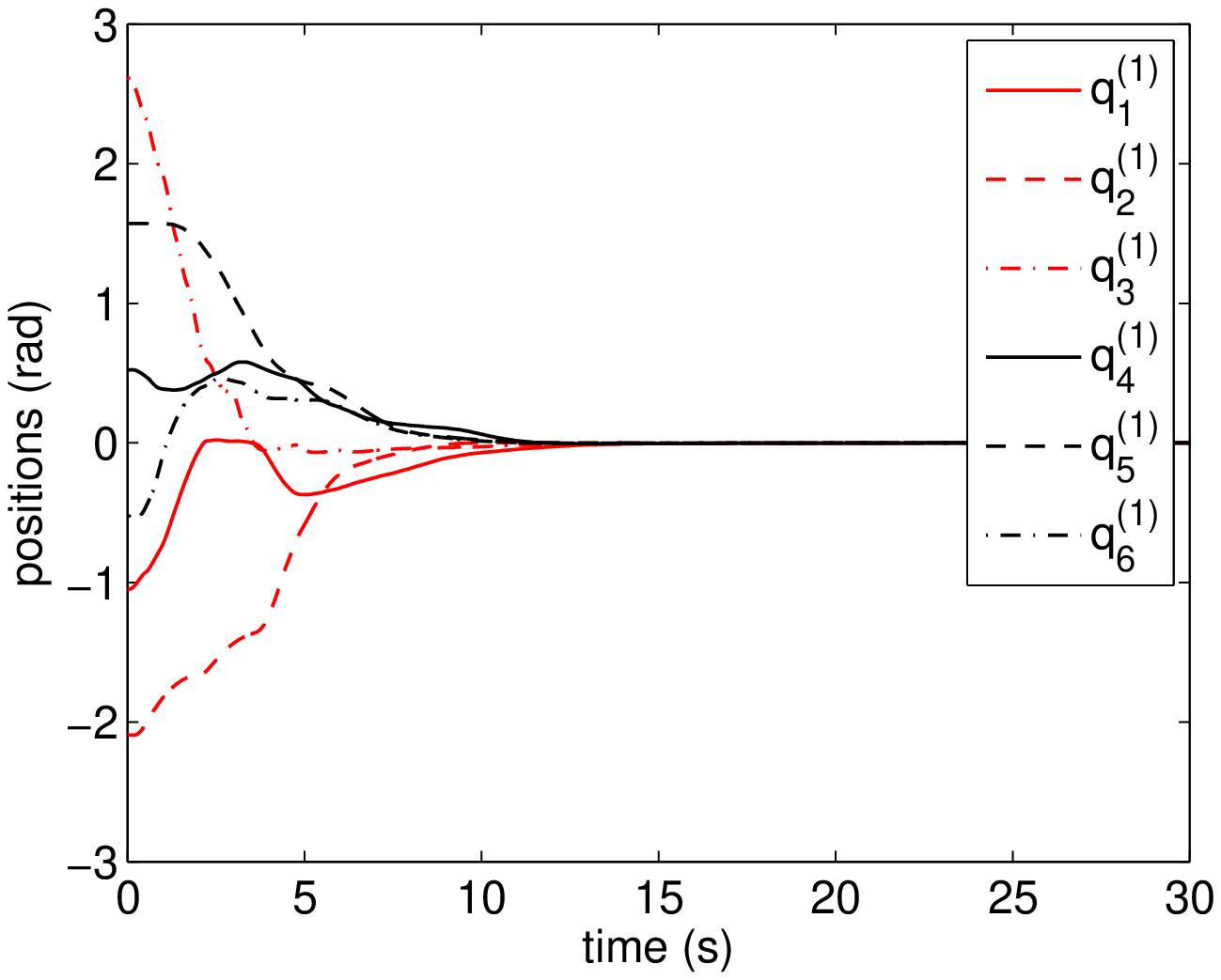}
\caption{Positions of the robots with the 3rd robot being manipulated by a human operator with $\lambda_{\mathcal M}=0$ (first coordinate).}
\end{minipage}%
\end{figure}

\begin{figure}
\centering
\begin{minipage}[t]{1.0\linewidth}
\centering
\includegraphics[width=3.2in]{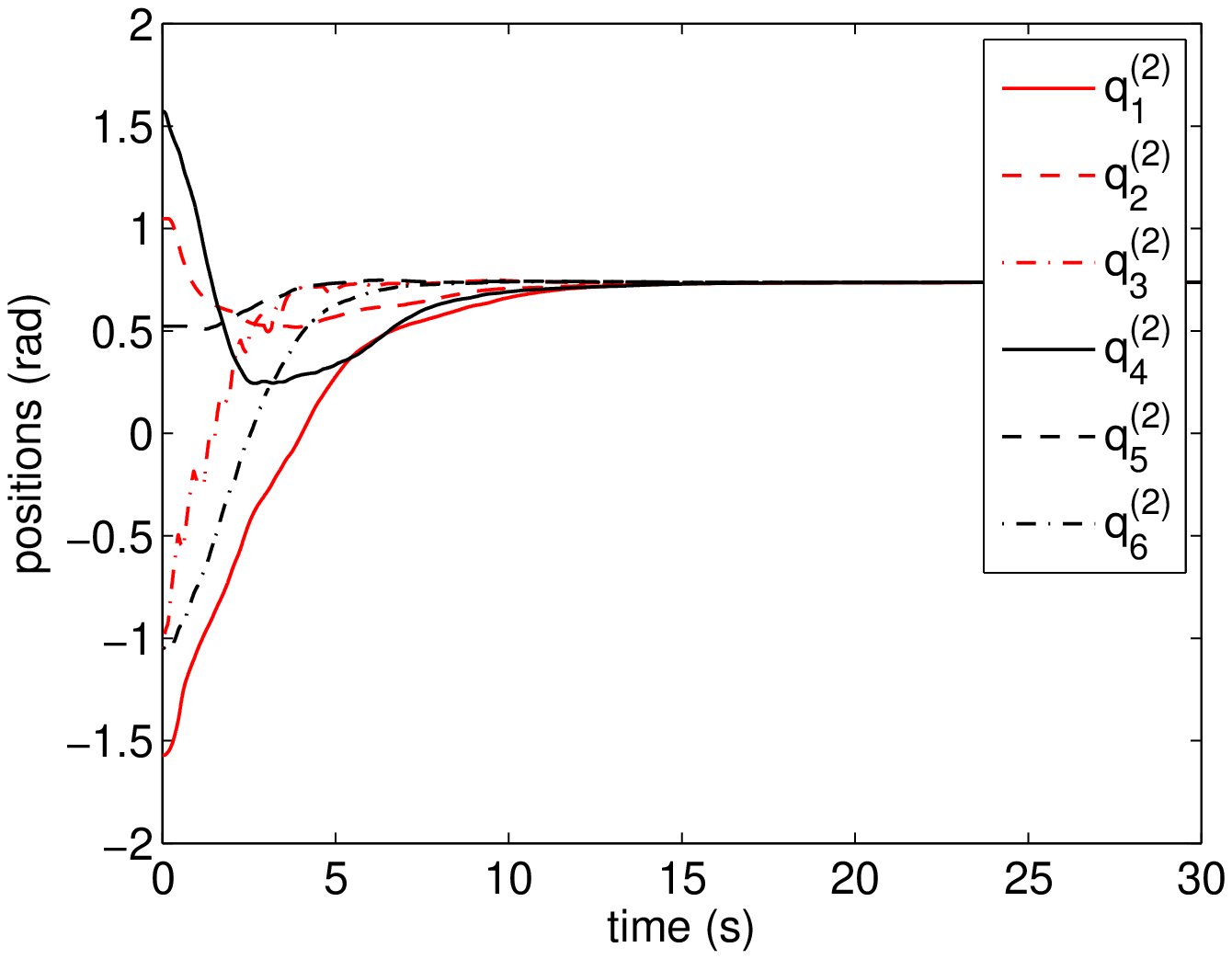}
\caption{Positions of the robots with the 3rd robot being manipulated by a human operator with $\lambda_{\mathcal M}=0$ (second coordinate).}
\end{minipage}%
\end{figure}

\begin{figure}
\centering
\begin{minipage}[t]{1.0\linewidth}
\centering
\includegraphics[width=3.2in]{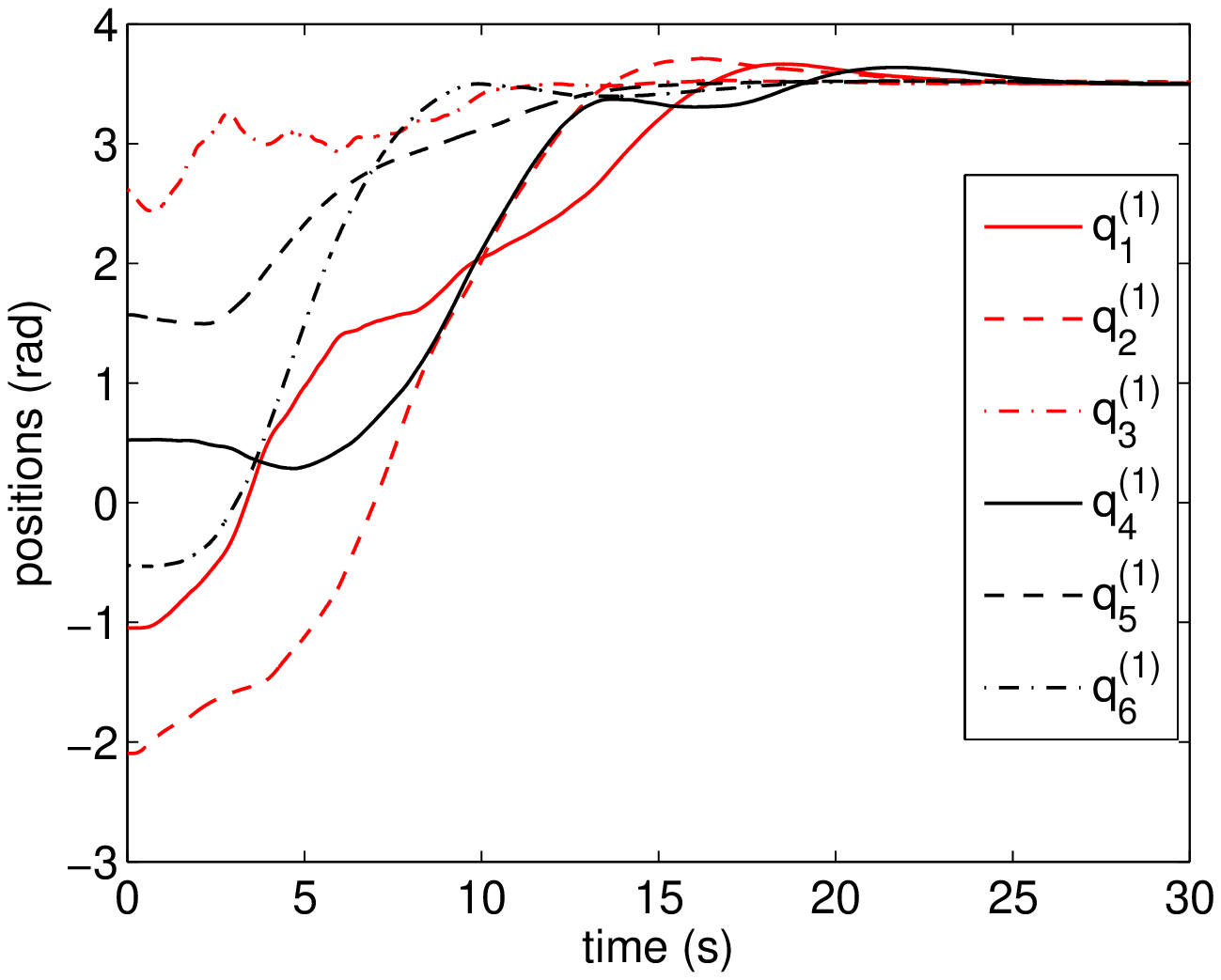}
\caption{Positions of the robots with the 3rd robot being manipulated by a human operator and with switching topology and communication delay (first coordinate).}
\end{minipage}%
\end{figure}

\begin{figure}
\centering
\begin{minipage}[t]{1.0\linewidth}
\centering
\includegraphics[width=3.2in]{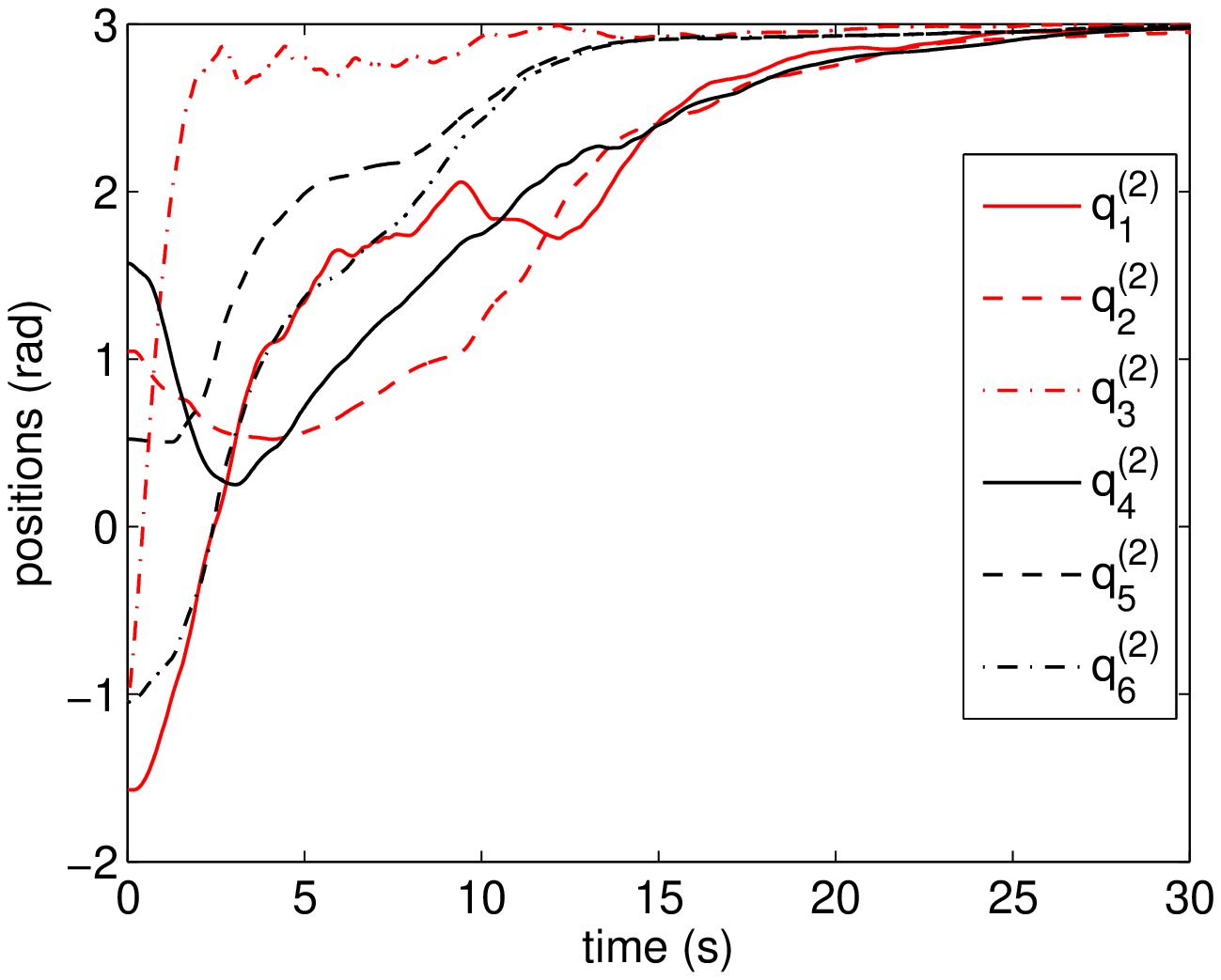}
\caption{Positions of the robots with the 3rd robot being manipulated by a human operator and with switching topology and communication delay (second coordinate).}
\end{minipage}%
\end{figure}

\subsection{Bilateral Teleoperation With Time-Varying Delay}

We now consider the case of bilateral teleoperation involving the first two robots used in Sec. VII-B and one acts as the master and the other acts as the slave. The context is quite the same except that the interaction topology between the two robots now becomes time-invariant. The controller parameters are chosen as $K=16.0 I_2$, $\Gamma_1=\Gamma_2=1.0I_3$, $\alpha=0.5$, $\lambda=2.0$, and $\lambda_{\mathcal M}=10.0$. The master robot (i.e., the first one) is manipulated by a human operator who exerts the standard PD control action as $\tau_1^\ast=-5.0\dot q_1-10.0(q_1-q_h)$ with $q_h=[3.5,3.0]^T$ being the desired position. The simulation results are shown in Fig. 12 and Fig. 13. For comparison, we perform another simulation with $\lambda_{\mathcal M}$ being reduced to $\lambda_{\mathcal M}=1.0$ and the simulation results are shown in Fig. 14 and Fig. 15. This shows the significance of the magnitude of the gain excluding that corresponding to the pure integral operation; even if in both the two cases, the infinite manipulability is achieved, the performance is, however, still quite different due to the fact that the infinite manipulability in the two cases has different increasing speeds with respect to the system operational frequency.

\begin{figure}
\centering
\begin{minipage}[t]{1.0\linewidth}
\centering
\includegraphics[width=3.2in]{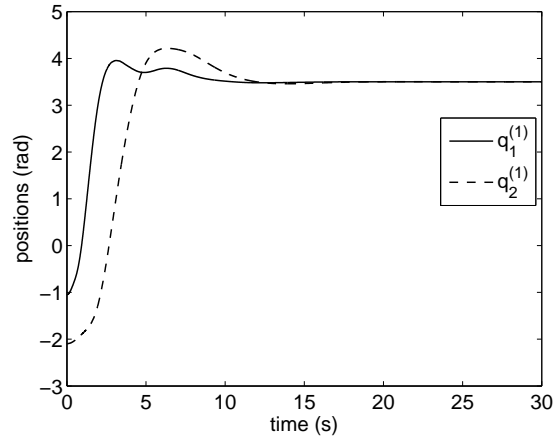}
\caption{Positions of the master and slave robots with $\lambda_{\mathcal M}=10.0$ (first coordinate).}
\end{minipage}%
\end{figure}

\begin{figure}
\centering
\begin{minipage}[t]{1.0\linewidth}
\centering
\includegraphics[width=3.2in]{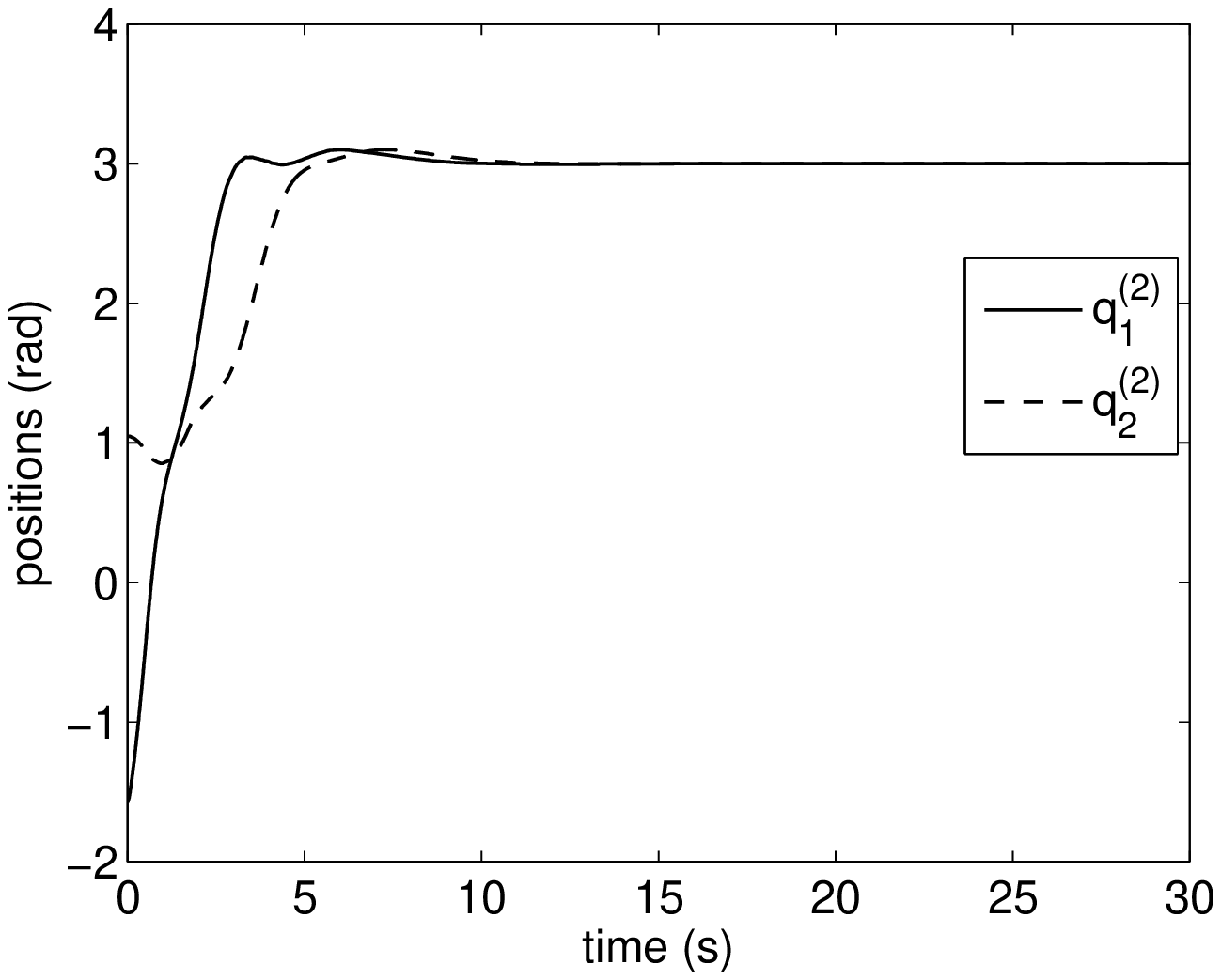}
\caption{Positions of the master and slave robots with $\lambda_{\mathcal M}=10.0$ (second coordinate).}
\end{minipage}%
\end{figure}

\begin{figure}
\centering
\begin{minipage}[t]{1.0\linewidth}
\centering
\includegraphics[width=3.2in]{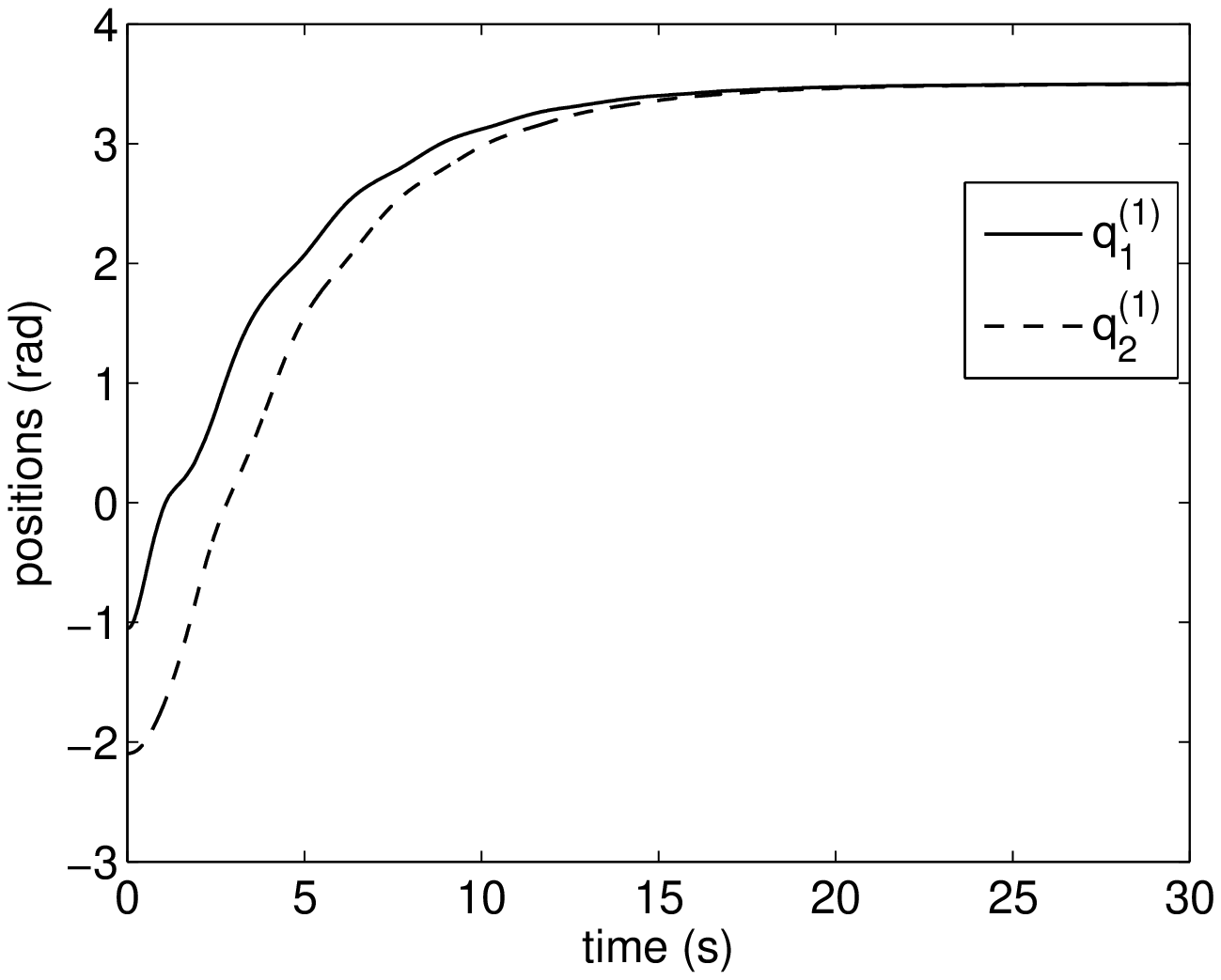}
\caption{Positions of the master and slave robots with $\lambda_{\mathcal M}=1.0$ (first coordinate).}
\end{minipage}%
\end{figure}

\begin{figure}
\centering
\begin{minipage}[t]{1.0\linewidth}
\centering
\includegraphics[width=3.2in]{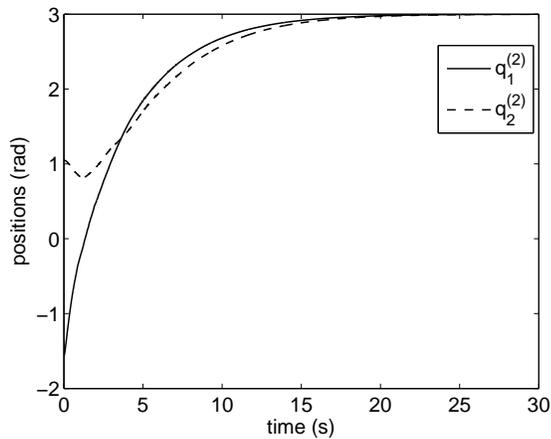}
\caption{Positions of the master and slave robots with $\lambda_{\mathcal M}=1.0$ (second coordinate).}
\end{minipage}%
\end{figure}

\section{Conclusion}

In this paper, we have systematically formulated the concept of infinite dynamical manipulability or simply infinite manipulability for dynamical systems and then investigated how a unified motivation based on this concept yields a systematic design paradigm for general interactive dynamical systems and interactive Lagrangian systems with parametric uncertainty and communication/sensing constraints. Specifically, the proposed design paradigm guarantees the infinite manipulability of the controlled Lagrangian systems with particularly strong robustness with respect to the interaction topology and time-varying communication delay. In addition, our result provides a solution to the longstanding benchmark problem of nonlinear bilateral teleoperation with arbitrary unknown time-varying communication delay, and in fact, our result gives the first delay-independent (independent of the time-varying delay) nonlinear adaptive teleoperation controller (to the best of our knowledge).

We would like to further discuss the connection between the physics of human-system interaction and mathematical properties of general functions, which becomes particularly prominent in the present work and shows some interesting features that might arouse our sense/adimiration of the delicate connection between the pure mathematics and physics (which has been historically witnessed for numerous times in various context). Our result can be considered as a contribution to this (probable)
historical truth/fact from the perspective of systems and control. Specifically, what attracts our attention in this study are those functions that are square-integrable yet not integral bounded.
As is shown in our main result, the insertion of a controlled square-integrable function that is not integral bounded (this function is generated by the closed-loop system) is crucial for ensuring both the easy manipulation of the system (this yields the infinite manipulability of the closed-loop system and consequently reduces the required amount of effort of the human operator) and the asymptotic position consensus (synchronization) among the Lagrangian systems. Square-integrability of functions often leads to the consequence that they converge to zero (for instance, if the functions are further uniformly continuous), which is well recognized in the field of systems and control. On the other hand, it is also well known that some of the square-integrable functions hold the possibility that their integrals with respect to time are unbounded. 
 Our study shows how such a property concerning general functions is systematically exploited in designing nonlinear controllers for interactive Lagrangian systems and associated with the gain properties of dynamical systems such as infinite manipulability and physical properties such as finite amount of energy.


%



\section*{Acknowledgment}

The author would like to thank Dr. Tiantian Jiang for the helpful comments on the paper.




\bibliographystyle{IEEEtran}
\bibliography{..//Reference_list_Wang}

\begin{thebibliography}{10}
\providecommand{\url}[1]{#1}
\csname url@samestyle\endcsname
\providecommand{\newblock}{\relax}
\providecommand{\bibinfo}[2]{#2}
\providecommand{\BIBentrySTDinterwordspacing}{\spaceskip=0pt\relax}
\providecommand{\BIBentryALTinterwordstretchfactor}{4}
\providecommand{\BIBentryALTinterwordspacing}{\spaceskip=\fontdimen2\font plus
\BIBentryALTinterwordstretchfactor\fontdimen3\font minus
  \fontdimen4\font\relax}
\providecommand{\BIBforeignlanguage}[2]{{%
\expandafter\ifx\csname l@#1\endcsname\relax
\typeout{** WARNING: IEEEtran.bst: No hyphenation pattern has been}%
\typeout{** loaded for the language `#1'. Using the pattern for}%
\typeout{** the default language instead.}%
\else
\language=\csname l@#1\endcsname
\fi
#2}}
\providecommand{\BIBdecl}{\relax}
\BIBdecl

\bibitem{Weiss1987_JRA}
L.~E. Weiss, A.~C. Sanderson, and C.~P. Neuman, ``Dynamic sensor-based control
  of robots with visual feedback,'' \emph{IEEE Journal of Robotics and
  Automation}, vol.~3, no.~5, pp. 404--417, Oct. 1987.

\bibitem{Asada1989_TRA}
H.~Asada and H.~Izumi, ``Automatic program generation from teaching data for
  the hybrid control of robots,'' \emph{IEEE Transactions on Robotics and
  Automation}, vol.~5, no.~2, pp. 166--173, Apr. 1989.

\bibitem{Ikeuchi1994_TRA}
K.~Ikeuchi and T.~Suehiro, ``Toward an assembly plan from observation {P}art
  {I}: Task recognition with polyhedral objects,'' \emph{IEEE Transactions on
  Robotics and Automation}, vol.~10, no.~3, pp. 368--385, Jun. 1994.

\bibitem{Craig2005_Book}
J.~J. Craig, \emph{Introduction to Robotics: Mechanics and Control},
  3rd~ed.\hskip 1em plus 0.5em minus 0.4em\relax Upper Saddle River, NJ:
  Prentice-Hall, 2005.

\bibitem{Sheridan1993_TRA}
T.~B. Sheridan, ``Space teleoperation through time delay: review and
  prognosis,'' \emph{IEEE Transactions on Robotics and Automation}, vol.~9,
  no.~5, pp. 592--606, Oct. 1993.

\bibitem{Hokayem2006_AUT}
P.~F. Hokayem and M.~W. Spong, ``Bilateral teleoperation: {A}n historical
  survey,'' \emph{Automatica}, vol.~42, no.~12, pp. 2035--2057, Dec. 2006.

\bibitem{Anderson1989_TAC}
R.~J. Anderson and M.~W. Spong, ``Bilateral control of teleoperators with time
  delay,'' \emph{IEEE Transactions on Automatic Control}, vol.~34, no.~5, pp.
  494--501, May 1989.

\bibitem{Niemeyer1991_JOE}
G.~Niemeyer and J.-J.~E. Slotine, ``Stable adaptive teleoperation,'' \emph{IEEE
  Journal of Oceanic Engineering}, vol.~16, no.~1, pp. 152--162, Jan. 1991.

\bibitem{Niemeyer2004_IJRR}
------, ``Telemanipulation with time delays,'' \emph{The International Journal
  of Robotics Research}, vol.~23, no.~9, pp. 873--890, Sep. 2004.

\bibitem{Nuno2011a_AUT}
E.~Nu{\~n}o, L.~Basa{\~n}ez, and R.~Ortega, ``Passivity-based control for
  bilateral teleoperation: A tutorial,'' \emph{Automatica}, vol.~47, no.~3, pp.
  485--495, Mar. 2011.

\bibitem{Lee2006_TRO}
D.~Lee and M.~W. Spong, ``Passive bilateral teleoperation with constant time
  delay,'' \emph{IEEE Transactions on Robotics}, vol.~22, no.~2, pp. 269--281,
  Apr. 2006.

\bibitem{Chopra2008_AUT}
N.~Chopra, M.~W. Spong, and R.~Lozano, ``Synchronization of bilateral
  teleoperators with time delay,'' \emph{Automatica}, vol.~44, no.~8, pp.
  2142--2148, Aug. 2008.

\bibitem{Nuno2009_IJRR}
E.~Nu{\~n}o, L.~Basa{\~n}ez, R.~Ortega, and M.~W. Spong, ``Position tracking
  for non-linear teleoperators with variable time delays,'' \emph{The
  International Journal of Robotics Research}, vol.~28, no.~7, pp. 895--910,
  Jul. 2009.

\bibitem{Liu2013_AUT}
Y.-C. Liu and N.~Chopra, ``Control of semi-autonomous teleoperation system with
  time delays,'' \emph{Automatica}, vol.~49, no.~6, pp. 1553--1565, Jun. 2013.

\bibitem{Wang2018_arXiv}
H.~Wang and Y.~Xie, ``Task-space consensus of networked robotic systems:
  Separation and manipulability,'' \emph{arXiv preprint arXiv:1702.06265},
  2018.

\bibitem{Nuno2011_TAC}
E.~Nu{\~n}o, R.~Ortega, L.~Basa{\~n}ez, and D.~Hill, ``Synchronization of
  networks of nonidentical {E}uler-{L}agrange systems with uncertain parameters
  and communication delays,'' \emph{IEEE Transactions on Automatic Control},
  vol.~56, no.~4, pp. 935--941, Apr. 2011.

\bibitem{Wang2014_TAC}
H.~Wang, ``Consensus of networked mechanical systems with communication delays:
  {A} unified framework,'' \emph{IEEE Transactions on Automatic Control},
  vol.~59, no.~6, pp. 1571--1576, Jun. 2014.

\bibitem{Abdessameud2014_TAC}
A.~Abdessameud, I.~G. Polushin, and A.~Tayebi, ``Synchronization of
  {L}agrangian systems with irregular communication delays,'' \emph{IEEE
  Transactions on Automatic Control}, vol.~59, no.~1, pp. 187--193, Jan. 2014.

\bibitem{Nuno2010_AUT}
E.~Nu{\~n}o, R.~Ortega, and L.~Basa{\~n}ez, ``An adaptive controller for
  nonlinear teleoperators,'' \emph{Automatica}, vol.~46, no.~1, pp. 155--159,
  Jan. 2010.

\bibitem{Nuno2017_IJACSP}
E.~Nu{\~n}o, C.~I. Aldana, and L.~Basa{\~n}ez, ``Task space consensus in
  networks of heterogeneous and uncertain robotic systems with variable
  time-delays,'' \emph{International Journal of Adaptive Control and Signal
  Processing}, vol.~31, no.~6, pp. 917--937, Jun. 2017.

\bibitem{Liu2014_SCL}
Y.~Liu, H.~Min, S.~Wang, Z.~Liu, and S.~Liao, ``Distributed adaptive consensus
  for multiple mechanical systems with switching topologies and time-varying
  delay,'' \emph{Systems \& Control Letters}, vol.~64, pp. 119--126, Feb. 2014.

\bibitem{Liu2015_JFI}
Y.-C. Liu, ``Distributed synchronization for heterogeneous robots with
  uncertain kinematics and dynamics under switching topologies,'' \emph{Journal
  of the Franklin Institute}, vol. 352, no.~9, pp. 3808--3826, Sep. 2015.

\bibitem{Chopra2008}
N.~Chopra and M.~W. Spong, ``Output synchronization of nonlinear systems with
  relative degree one,'' in \emph{Recent Advances in Learning and Control},
  V.~D. Blondel, S.~P. Boyd, and H.~Kimura, Eds.\hskip 1em plus 0.5em minus
  0.4em\relax London, U.K.: Springer-Verlag, 2008, pp. 51--64.

\bibitem{Munz2011b_TAC}
U.~M{\"u}nz, A.~Papachristodoulou, and F.~Allg{\"o}wer, ``Consensus in
  multi-agent systems with coupling delays and switching topology,'' \emph{IEEE
  Transactions on Automatic Control}, vol.~56, no.~12, pp. 2976--2982, Dec.
  2011.

\bibitem{Rezaee2017_IJRNC}
H.~Rezaee and F.~Abdollahi, ``Adaptive stationary consensus protocol for a
  class of high-order nonlinear multiagent systems with jointly connected
  topologies,'' \emph{International Journal of Robust and Nonlinear Control},
  vol.~27, no.~9, pp. 1677--1689, Jun. 2017.

\bibitem{Liu2014_JFI}
Y.~Liu, H.~Min, S.~Wang, L.~Ma, and Z.~Liu, ``Consensus for multiple
  heterogeneous {E}uler--{L}agrange systems with time-delay and jointly
  connected topologies,'' \emph{Journal of the Franklin Institute}, vol. 351,
  no.~6, pp. 3351--3363, Jun. 2014.

\bibitem{Wang2017_CAC}
H.~Wang, ``Dynamic feedback for consensus of networked {L}agrangian systems
  with switching topology,'' in \emph{China Automation Congress}, Ji'nan,
  China, 2017, pp. 1340--1345.

\bibitem{Wang2017_AUTSubmitted}
------, ``Integral-cascade framework for consensus of networked {L}agrangian
  systems,'' submitted to \emph{{A}utomatica}.

\bibitem{Chopra2003_ACC}
N.~Chopra, M.~W. Spong, S.~Hirche, and M.~Buss, ``Bilateral teleoperation over
  the internet: the time varying delay problem,'' in \emph{Proceedings of the
  American Control Conference}, Denver, CO, USA, 2003, pp. 155--160.

\bibitem{Slotine1987_IJRR}
J.-J.~E. Slotine and W.~Li, ``On the adaptive control of robot manipulators,''
  \emph{The International Journal of Robotics Research}, vol.~6, no.~3, pp.
  49--59, Sep. 1987.

\bibitem{Slotine1991_Book}
------, \emph{Applied Nonlinear Control}.\hskip 1em plus 0.5em minus
  0.4em\relax Englewood Cliffs, NJ: Prentice-Hall, 1991.

\bibitem{Spong2006_Book}
M.~W. Spong, S.~Hutchinson, and M.~Vidyasagar, \emph{Robot Modeling and
  Control}.\hskip 1em plus 0.5em minus 0.4em\relax New York: Wiley, 2006.

\bibitem{Niemeyer1998_ICRA}
G.~Niemeyer and J.-J.~E. Slotine, ``Towards force-reflecting teleoperation over
  the internet,'' in \emph{Proceedings of the IEEE International Conference on
  Robotics and Automation}, Leuven, Belgium, 1998, pp. 1909--1915.

\bibitem{Desoer1975_Book}
C.~A. Desoer and M.~Vidyasagar, \emph{Feedback Systems: Input-Output
  Properties}.\hskip 1em plus 0.5em minus 0.4em\relax New York: Academic Press,
  1975.

\bibitem{Ioannou1996_Book}
P.~A. Ioannou and J.~Sun, \emph{Robust Adaptive Control}.\hskip 1em plus 0.5em
  minus 0.4em\relax Englewood Cliffs, NJ: Prentice-Hall, 1996.

\bibitem{Schaft2000_Book}
A.~J. van~der Schaft, \emph{${\mathcal L}_2$-Gain and Passivity Techniques in
  Nonlinear Control}, 2nd~ed.\hskip 1em plus 0.5em minus 0.4em\relax London:
  Springer-Verlag, 2000.

\bibitem{Khalil2002_Book}
H.~K. Khalil, \emph{Nonlinear Systems}, 3rd~ed.\hskip 1em plus 0.5em minus
  0.4em\relax Upper Saddle River, NJ: Prentice-Hall, 2002.

\bibitem{Vidyasagar1993_Book}
M.~Vidyasagar, \emph{Nonlinear Systems Analysis}, 2nd~ed.\hskip 1em plus 0.5em
  minus 0.4em\relax Englewood Cliffs, NJ: Prentice-Hall, 1993.

\bibitem{Rotea1993_AUT}
M.~A. Rotea, ``The generalized ${H}_2$ control problem,'' \emph{Automatica},
  vol.~29, no.~2, pp. 373--385, Mar. 1993.

\bibitem{Sontag1998_SCL}
E.~D. Sontag, ``Comments on integral variants of {ISS},'' \emph{Systems \&
  Control Letters}, vol.~34, no. 1-2, pp. 93--100, May 1998.

\bibitem{Angeli2000_TAC}
D.~Angeli, E.~D. Sontag, and Y.~Wang, ``A characterization of integral
  input-to-state stability,'' \emph{IEEE Transactions on Automatic Control},
  vol.~45, no.~6, pp. 1082--1097, Jun. 2000.

\bibitem{Slotine1989_Aut}
J.-J.~E. Slotine and W.~Li, ``Composite adaptive control of robot
  manipulators,'' \emph{Automatica}, vol.~25, no.~4, pp. 509--519, Jul. 1989.

\bibitem{Wang2017_TAC}
H.~Wang, ``Adaptive control of robot manipulators with uncertain kinematics and
  dynamics,'' \emph{IEEE Transactions on Automatic Control}, vol.~62, no.~2,
  pp. 948--954, Feb. 2017.

\bibitem{Ortega1989_AUT}
R.~Ortega and M.~W. Spong, ``Adaptive motion control of rigid robots: {A}
  tutorial,'' \emph{Automatica}, vol.~25, no.~6, pp. 877--888, Nov. 1989.

\bibitem{Olfati-Saber2004_TAC}
R.~Olfati-Saber and R.~M. Murray, ``Consensus problems in networks of agents
  with switching topology and time-delays,'' \emph{IEEE Transactions on
  Automatic Control}, vol.~49, no.~9, pp. 1520--1533, Sep. 2004.

\bibitem{Ren2005_TAC}
W.~Ren and R.~W. Beard, ``Consensus seeking in multiagent systems under
  dynamically changing interaction topologies,'' \emph{IEEE Transactions on
  Automatic Control}, vol.~50, no.~5, pp. 655--661, May 2005.

\bibitem{Ren2008_Book}
------, \emph{Distributed Consensus in Multi-Vehicle Cooperative
  Control}.\hskip 1em plus 0.5em minus 0.4em\relax London, U.K.:
  Springer-Verlag, 2008.

\bibitem{Chopra2006}
N.~Chopra and M.~W. Spong, ``Passivity-based control of multi-agent systems,''
  in \emph{Advances in Robot Control: From Everyday Physics to Human-Like
  Movements}, S.~Kawamura and M.~Svinin, Eds.\hskip 1em plus 0.5em minus
  0.4em\relax Berlin, Germany: Springer-Verlag, 2006, pp. 107--134.

\bibitem{Brewer1978_TCS}
J.~W. Brewer, ``Kronecker products and matrix caculus in system theory,''
  \emph{IEEE Transactions on Circuits and Systems}, vol. CAS-25, no.~9, pp.
  772--781, Sep. 1978.

\bibitem{Lee2007_TAC}
D.~Lee and M.~W. Spong, ``Stable flocking of multiple inertial agents on
  balanced graphs,'' \emph{IEEE Transactions on Automatic Control}, vol.~52,
  no.~8, pp. 1469--1475, Aug. 2007.

\bibitem{Jiang2009_IET}
H.~B. Jiang, ``Hybrid adaptive and impulsive synchronisation of uncertain
  complex dynamical networks by the generalised {B}arbalat's lemma,'' \emph{IET
  Control Theory \& Applications}, vol.~3, no.~10, pp. 1330--1340, Oct. 2009.

\bibitem{Yokokohji1999_IROS}
Y.~Yokokohji, T.~Imaida, and T.~Yoshikawa, ``Bilateral teleoperation under
  time-varying communication delay,'' in \emph{Proceedings of the IEEE/RSJ
  International Conference on Intelligent Robots and Systems}, Kyongju, South
  Korea, 1999, pp. 1854--1859.

\bibitem{Munir2002_TMECH}
S.~Munir and W.~J. Book, ``Internet-based teleoperation using wave variables
  with prediction,'' \emph{IEEE/ASME Transactions on Mechatronics}, vol.~7,
  no.~2, pp. 124--133, Jun. 2002.

\bibitem{Ching2006_JDSMC}
H.~Ching and W.~J. Book, ``Internet-based bilateral teleoperation based on wave
  variable with adaptive predictor and direct drift control,'' \emph{Journal of
  Dynamic Systems, Measurement, and Control}, vol. 128, no.~1, pp. 86--93, Mar.
  2006.

\bibitem{Chopra2008_TCST}
N.~Chopra, P.~Berestesky, and M.~W. Spong, ``Bilateral teleoperation over
  unreliable communication networks,'' \emph{IEEE Transactions on Control
  Systems Technology}, vol.~16, no.~2, pp. 304--313, Mar. 2008.

\bibitem{Chopra2006_TRO}
N.~Chopra, M.~W. Spong, R.~Ortega, and N.~E. Barabanov, ``On tracking
  performance in bilateral teleoperation,'' \emph{IEEE Transactions on
  Robotics}, vol.~22, no.~4, pp. 861--866, Aug. 2006.

\bibitem{Polushin2006_TCybernetics}
I.~G. Polushin, P.~X. Liu, and C.-H. Lung, ``A control scheme for stable
  force-reflecting teleoperation over ip networks,'' \emph{IEEE Transactions on
  Systems, Man, and Cybernetics--Part B: Cybernetics}, vol.~36, no.~4, pp.
  930--939, Aug. 2006.

\bibitem{Lee2010_TRO_PassiveSetPosition}
D.~Lee and K.~Huang, ``Passive-set-position-modulation framework for
  interactive robotic systems,'' \emph{IEEE Transactions on Robotics}, vol.~26,
  no.~2, pp. 354--369, Apr. 2010.

\bibitem{Liu2015_TMECH}
Y.-C. Liu and M.-H. Khong, ``Adaptive control for nonlinear teleoperators with
  uncertain kinematics and dynamics,'' \emph{IEEE/ASME Transactions on
  Mechatronics}, vol.~20, no.~5, pp. 2550--2562, Oct. 2015.

\bibitem{Polushin2015_TMECH}
I.~G. Polushin, A.~Takhmar, and R.~V. Patel, ``Projection-based
  force-reflection algorithms with frequency separation for bilateral
  teleoperation,'' \emph{IEEE/ASME Transactions on Mechatronics}, vol.~20,
  no.~1, pp. 143--154, Feb. 2015.

\bibitem{Nuno2018_IJRNC}
E.~Nu{\~n}o, M.~Arteaga-P{\'e}rez, and G.~Espinosa-P{\'e}rez, ``Control of
  bilateral teleoperators with time delays using only position measurements,''
  \emph{International Journal of Robust and Nonlinear Control}, vol.~28, no.~3,
  pp. 808--824, Feb. 2018.

\bibitem{Hua2017_TAC}
C.-C. Hua, X.~Yang, J.~Yan, and X.-P. Guan, ``On exploring the domain of
  attraction for bilateral teleoperator subject to interval delay and saturated
  {P}+d control scheme,'' \emph{IEEE Transactions on Automatic Control},
  vol.~62, no.~6, pp. 2923--2928, Jun. 2017.

\end{thebibliography}

%
%
%

%







\end{document}